\documentclass[
12pt, 
english, 
singlespacing, 
headsepline, 
chapterinoneline, 
]{MastersDoctoralThesis} 

\usepackage[utf8]{inputenc} 
\usepackage[T1]{fontenc} 

\usepackage{fourier} 

\usepackage{amsthm,amsmath,latexsym,amssymb,amsfonts,amsbsy}

\usepackage{tikz} 
\usepackage{pgfplots}
\let\oldsqrt\sqrt

   \def\sqrt{\mathpalette\DHLhksqrt}
   \def\DHLhksqrt#1#2{%
   \setbox0=\hbox{$#1\oldsqrt{#2\,}$}\dimen0=\ht0
   \advance\dimen0-0.2\ht0
   \setbox2=\hbox{\vrule height\ht0 depth -\dimen0}%
   {\box0\lower0.4pt\box2}} 

\usepackage[autostyle=true]{csquotes} 
\newcommand{\HRulegrossa}{\rule{\linewidth}{1.2mm}}

\thesistitle{\sc Black hole energy in Lovelock Unique Vacuum theories} 
\supervisor{Dr. Rodrigo \textsc{Olea}} 
\examiner{Dr. Olivera Miskovic (PUCV)\\Dr. Rodrigo Aros (UNAB)}
\degree{Master of science} 
\author{Gabriel \textsc{Arenas-Henriquez}} 
\addresses{} 

\subject{Physics science} 
\university{\href{http://www.unab.cl}{Universidad Andres Bello}} 
\department{\href{http://fisica.unab.cl}{Physics Department}} 
\group{\href{}{Faculty of Exact Sciences}} 
\faculty{\href{http://facultades.unab.cl/cienciasexactas/}{Faculty of Exact Sciences}} 

\AtBeginDocument{
\hypersetup{pdftitle=\ttitle} 
\hypersetup{pdfauthor=\authorname} 
\hypersetup{pdfkeywords=\keywordnames} 
\hypersetup{colorlinks=false}
\hypersetup{urlcolor=red}
}

\begin{document}

\frontmatter 

\pagestyle{plain} 


\begin{titlepage}
 
\begin{center}
\includegraphics[width=0.3\textwidth]{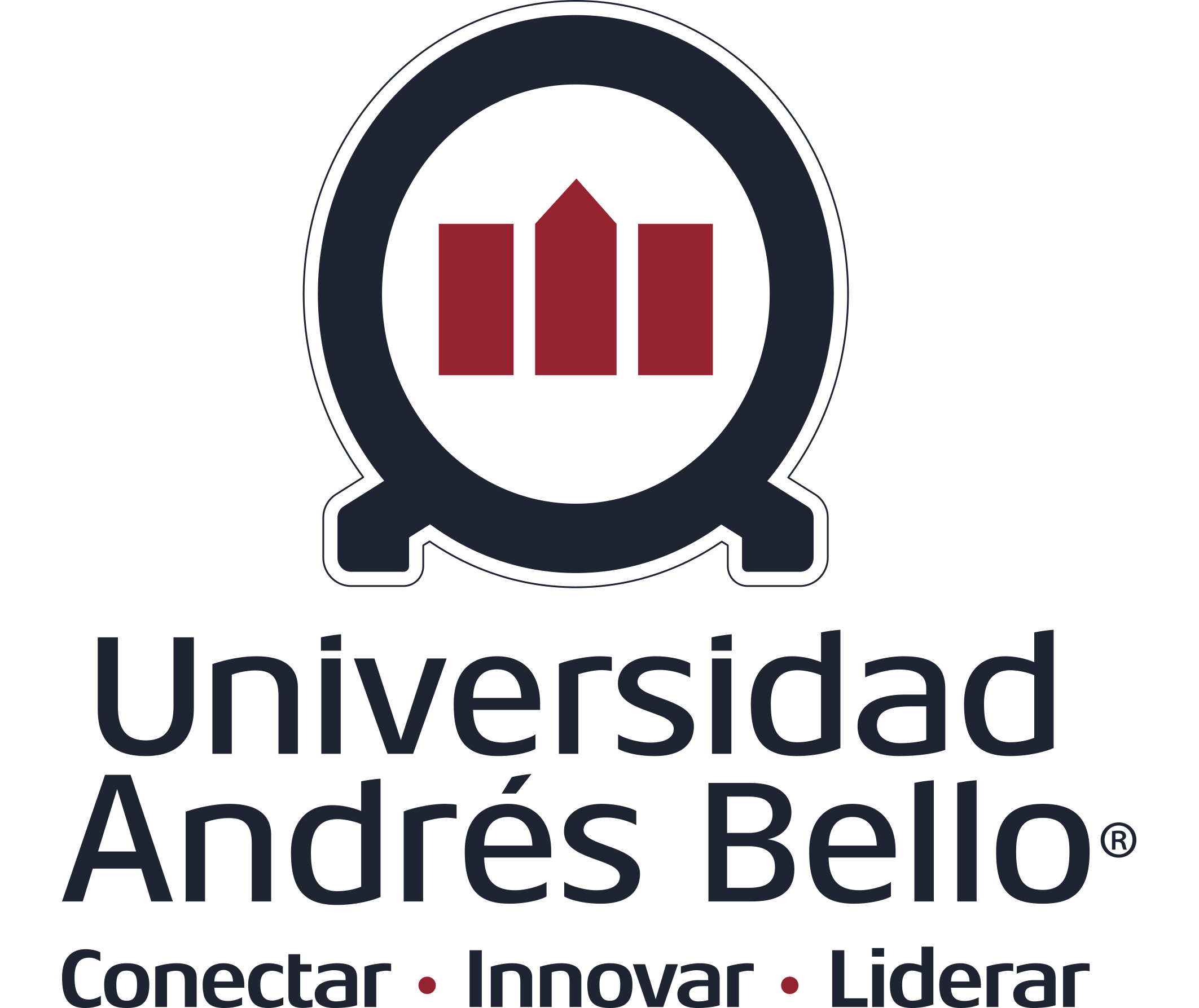}

\vspace*{.06\textheight}
{\scshape\LARGE \univname\par}\vspace{1.5cm} 
\textsc{\Large Master Thesis}\\[0.5cm] 

\HRulegrossa \\
\HRule \\[0.4cm] 
{\huge \bfseries \ttitle\par}\vspace{0.4cm} 
\HRule \\ 
 \HRulegrossa \\[1.5 cm]
 
\begin{minipage}[t]{0.4\textwidth}
\begin{flushleft} \large
\emph{Author:}\\
\href{http://inspirehep.net/author/profile/G.Arenas.Henriquez.1}{\authorname} 
\end{flushleft}
\end{minipage}
\begin{minipage}[t]{0.5\textwidth}
\begin{flushright} \large
\emph{Supervisor:} \\
\href{http://www.hologravity.com}{\supname}\\
\end{flushright}
\end{minipage}\\[2cm]
\vfill

\emph{Jury Members:}\\
\examname
\vspace{2mm}

\large \textit{A thesis submitted in fulfillment of the requirements\\ for the degree of \degreename}\\[0.3cm] 
\textit{in the}\\[0.4cm]
\groupname\\\deptname\\[2cm] 

\end{center}
\end{titlepage}


\cleardoublepage


\vspace*{0.2\textheight}

\noindent\enquote{\itshape Confusion will be my epitaph.}\bigbreak

\hfill Peter Sinfield.


\begin{abstract}
\addchaptertocentry{\abstractname} 
In this thesis, we explore the possibility to define black hole mass in terms of the Weyl tensor for the entire family of Lovelock-AdS gravity theories. The level of degeneracy $k$ of the corresponding vacuum fixes the number of curvatures that should appear in the energy formula.  Therefore, the charge expression which is a polynomial of maximal degree in the curvature, can be consistently truncated to an order $k$ in the Weyl tensor. In particular,
 for the maximally degenerate case in odd dimensions (Chern-Simons AdS) the expression identically vanishes and the mass must come from the formula that, in the other cases, produces the vacuum energy.
\end{abstract}


\begin{acknowledgements}
\addchaptertocentry{\acknowledgementname} 
First of all, I would like to thank to my supervisor, Rodrigo Olea for the innumerable discussions about physics, life and music. Thank you for all the support and guidance. Also, I would like to thank to whom I consider my second supervisor, Olivera Miskovic. I am deeply grateful for reading careful this thesis, for all her comments and also, for all the tricks that she taught me during the time that we have worked together. Both of you have stimulated my interest and passion for theoretical physics and even more than that. 

\indent

Secondly, I want to thank to all the people that have spent time in the \textit{HEP} group of UNAB for all the discussions and seminars that have feed my interest even beyond high-energy physics. In particular, Giorgos Anastasiou, Ignacio Araya, Cesar Arias, Israel Ramirez Krause, Omar Rodriguez-Tzompantzi, Yihao Yin, Yerko Novoa,
Pablo Guilleminot, Gustavo Valdivia, Javier Moreno, Ivan Cascan, Per Sundell, Fabian Caro, Matias Pinto and Rodrigo Aros. Special mentions to Felipe Diaz and David Rivera-Betancour with whom I have shared the whole process, that started with a rigorous training on a summer of three years ago. We all can say now that we have learned a few things about physics, hikes and life.

\indent 

I am also grateful to professor Robert B. Mann for all his detailed comments, corrections and kind collaboration.

\indent 

I want to thank to Veronica Gaete, Alexandra Ovalle, Tomas Frias, Camilo Weinberger, Rene Zuñiga and Veronica Ovalle, for all the friendship and moments that we have shared. Thank you all. 

\indent

Finally, I would like to express my deepest feelings of gratitude to my whole family and, in particular, to my parents, sister and grandpa for their unfailing love and support of all my decisions. Without them, nothing of this would be possible. 

\indent

This work was supported in part by the Grants FONDECYT No 1170765 and UNAB M.Sc. Scholarship.
\end{acknowledgements}


\tableofcontents 

\dedicatory{Dedicated to my grandma,\\
            J. Lopez} 


\mainmatter 

\pagestyle{thesis} 



\chapter{Introduction} 

\label{Intro} 


\newcommand{\keyword}[1]{\textbf{#1}}
\newcommand{\tabhead}[1]{\textbf{#1}}
\newcommand{\code}[1]{\texttt{#1}}
\newcommand{\file}[1]{\texttt{\bfseries#1}}
\newcommand{\option}[1]{\texttt{\itshape#1}}


Black hole thermodynamics, introduced in early seventies  \cite{Bekenstein,bardeen1973,hawking1975,Hawking1976}, links classical aspects of gravity with underlying properties of a possible quantum theory of gravity. It has stimulated the development of plenty of techniques in order to analyze the behavior of black holes and obtain relevant information about the quantum aspects of them. One of the most remarkable result is the area law for the entropy of a black hole (in fundamental units)
\begin{equation}
 S_{BH}=\frac{A}{4}\,
\end{equation}
where $A$ corresponds to the area of the black hole horizon.
Inspired by this result, 't Hooft and Susskind proposed the Holographic Principle \cite{tHooft:1993dmi, Susskind:1994vu} where the information about the (quantum) gravity states is encoded in a lower-dimensional theory at the boundary. A concrete realization of this idea was conjectured by Maldacena \cite{Maldacena:1997re}, improved by Witten \cite{Witten:1998qj} and Gubser et al. {Gubser:1998bc}, in the so-called Anti-de Sitter/Conformal Field Theory (AdS/CFT) correspondence.  The duality suggest that there is a one-to-one relation between correlators in $\mathcal{N}=4$ Super Yang-Mills theory in the large \textit{N} limit and type IIB $10$-dimensional supergravity (SUGRA) on $AdS_{5}\times S^{5}$ background.  A precise holographic dictionary has been formulated \cite{Witten:1998zw,Balasubramanian:1999re,Henningson:1998gx,Aharony:1999ti} producing relevant results. A remarkable example was derived by Policastro, Son and Starinets \cite{Policastro:2001yc} where they computed the shear viscosity of quark-gluon plasma using a non-extremal black three-brane as holographic dual. 
\indent

Nevertheless, holographic properties of higher-curvature gravity theories are still unclear. That is the case of Lovelock-AdS gravity theories \cite{Lovelock:1971yv} which arises naturally from the low energy limit of string theory \cite{PhysRevLett.55.2656}. The rich AdS vacua structure and similar properties to Einstein AdS gravity has motivated the study of this family of theories for holographic applications. 
\indent

In particular, Lovelock Unique Vacuum theories \cite{Crisostomo:2000bb,Kastor:2006vw} are obtained by a proper choice of coupling constants in the Lovelock lagrangian. They possesses a unique AdS vacuum and thereby, they are degenerate theories. It has been shown that linearization around a fixed background in such theories breaks down and thus, it is not possible to properly identify the graviton as a propagating field \cite{Camanho:2009hu,Camanho:2010ru}. As a matter of fact, the holographic implications of the degeneracy are not well understood and further insights of its role is needed.   \\
In this thesis, we focus on the definition conserved charges of Lovelock-AdS theories and the explicit connection of them with the degeneracy of the corresponding \textit{branches}. It is organized as follows:
\indent

 In \autoref{AdS/CFT}, we review some basics concepts in anti-de Sitter spacetimes and their connection to gauge/gravity correspondence. We will also present, in a summarized way, the standard recipe for holographic renormalization and obtaining correlation functions. This chapter is in part based on \cite{Penedones:2016voo,Skenderis:2002wp}. 
\indent 

 In \autoref{Lovelock}, we introduce a generic Lovelock Lagrangian and we particularize the theory in order to contain AdS solutions. We also obtain the degeneracy condition of order $k$ for a vacuum solution of the equation of motion. Then, we analyze the asymptotic behavior for static black holes in Lovelock-AdS gravity, generalized for any level of degeneracy. The explicit falloff of the solution will allow to evaluate the on-shell Weyl tensor and compare it with the AdS curvature. \\
 \indent
 In \autoref{NoetherTheorem}, we introduce the Kounterterm method and the Noether theorem for diffeomorphism-invariant theories. Using the results for Lovelock-AdS action, we express the conserved charges in terms of a power of the Weyl tensor, arriving at the final mass formula \cite{Arenas-Henriquez:2019rph}. 
 \indent
 
 Finally, in \autoref{conclusions}, we provide some concluding remarks and we foresee future directions of our work.
 

\chapter{Asymptotically AdS spacetimes and gauge/gravity duality} 

\label{AdS/CFT} 


General Relativity was formulated by Albert Einstein and published in 1915 \cite{Einstein:1916vd}. The dynamics of the theory is described by the equation
\begin{equation}
\mathcal{E}_{\mu \nu}=R_{\mu \nu} -\frac{1}{2}Rg_{\mu \nu}+\Lambda g_{\mu \nu}=8\pi G T_{\mu \nu}\,, \label{einstein}
\end{equation}
where $\Lambda$ is the cosmological constant, $G$ is the Newton constant and $T_{\mu \nu}$ the energy-momentum tensor of the matter fields. Basically, the equation shows how the matter affects the geometry of the spacetimes and how the geometry backreacts to affect the way matter is moving. However, the dynamic content of the theory is not enough for an entire description of it. David Hilbert, in parallel to Einstein, constructed the action principle for General Relativity, of what today is called Einstein-Hilbert action. In a $D$-dimensional manifold $\mathcal{M}$ without considering matter 
\begin{equation}
    I_{EH}=\frac{1}{16\pi G}\int \limits_{\mathcal{M}}^{}d^{D}x\sqrt{-g}\left(R-2\Lambda\right) \,,  \label{EH}
\end{equation}
with the volume element $\sqrt{-g}$ where $g$ is the determinant of the metric $g_{\mu \nu}$.  For arbitrary variations of the action with respect to the field, i.e., the metric will produce the equations of motion (\ref{einstein}). The solutions of these equations have been extensively studied being Karl Schwarzschild in 1916 the first in giving a solution: the Schwarzschild black hole. Just by considering a spherically symmetric stellar (or massive) object, Schwarzschild proved that symmetries on the spacetime are enough to find an explicit solution that satisfies the Einstein's equations. 
Nevertheless, one could ask if there exists a way to construct a vacuum configuration with the maximum number of isometries without being trivial. The easier answer would be the flat spacetime and which is solution to Einstein's theory by simple consistency with special relativity.
At least if we consider a Minkowski spacetime in $D$-dimensions, where the cosmological constant vanishes, and we count the number of isometries, we are going to find that there are $D(D+1)/2$ Killing vectors $\xi^{\mu}$ from where we can distinguish $D(D-1)/2$ boosts and $D$ quasitranslations. Now, we could try to find a new solution with the same number of Killing vectors but including cosmological constant. The answer will be a constant curvature spacetime proportional to the cosmological constant
\begin{equation}
  R_{\mu \nu \alpha \beta}= \frac{\Lambda}{D(D-1)}\left(g_{\mu \alpha} g_{\nu \beta} -g_{\mu \beta}g_{\nu \alpha}\right). \label{Riem}
\end{equation}
It is obvious from here that the value of $\Lambda$ defines the sign of the curvature for the spacetime. In fact, the cosmological constant is related to the characteristic length of the radius $\ell$ of the space as
\begin{equation}
  \Lambda=\pm \frac{(D-1)(D-2)}{2\ell^{2}}   \,.
\end{equation}
A positive cosmological constant and, therefore, positive curvature corresponds to \textit{de Sitter} (dS) spacetime named in honour to Willem de Sitter who discussed the possibility to describe an expanding universe. 
\indent

On the other hand, a negative cosmological constant and, thus, negative curvature defines what is called an \textit{anti-de Sitter} (AdS) spacetime.
\indent

Even though observational results suggest that our universe has a small positive cosmological constant as dS spacetime, the interest of many theoretical physicist have focused on AdS spacetimes by the appearance of gauge/gravity duality.\\
Following this line,  we are going to review few aspects of AdS spacetimes, their conformal structure and the link to AdS/CFT correspondence.

\section{AAdS spacetimes}
The global anti-de Sitter spacetime in $D=d+1$ dimensions and Lorentzian signature can be defined as the following quadratic form embedded in $(D+1)$-dimensional flat space with signature $(-,+,\cdots,+,-)$ which is invariant under the action of the group $SO(d,2)$
\begin{equation}
     -(x^{0})^{2}+(x^{1})^{2}+\dots +(x^{d})^{2} -(x^{d+1})^{2}=-\ell^{2}\,. \label{constraint}
\end{equation}
The corresponding line element, written in the embedding $\mathbb{R}^{d+2}$
\begin{equation}
    ds^{2}= -(dx^{0})^{2}+(dx^{1})^{2}+\dots +(dx^{d})^{2} -(dx^{d+1})^{2}\,.
\end{equation}
The quadratic form, subjected to the constraint (\ref{constraint}), can be rewritten in the local coordinates. For instance, there are two particular choices useful for holographic applications. Let us start with global coordinate patch  which covers the entire spacetime and it is given by 
\begin{eqnarray}
x^{0}&=&\ell \cos (t)\cosh (\rho) \nonumber \\
x^{i}&=&\ell \sinh (\rho)\Omega ^{i} \\
x^{d+1}&=& -\ell \sin (t) \cosh (\rho) \nonumber
\end{eqnarray}
where $i=1,\dots,d-1$, $t\in \mathbb{R}$ and $\Omega^{i}$ parametrizes a unit $S^{d-1}$. Thus, the metric becomes
\begin{equation}
ds^{2}=\ell ^{2}\left(-\cosh^{2} (\rho)dt^{2}+d\rho^{2}+\sinh^{2}(\rho)d\Omega_{d-1}^{2}\right)\,.
\end{equation}
A simple change of coordinates $\tan(\theta)=\sinh(\rho)$ such that $\theta \in [0,\frac{\pi}{2}]$ yields
\begin{equation}
 ds^{2}=\frac{\ell ^{2}}{cos^{2}(\theta)}\left(-dt^{2}+d\theta^{2}+\sin^{2}(\theta)d\Omega_{d-1}^{2}\right) \,. 
\end{equation}
Evidently, the last metric is conformally equivalent to a solid cylinder whose boundary is located at $\theta=\frac{\pi}{2}$. Nevertheless, the conformal factor in front induces a pole of order two (singularity), on the asymptotic boundary, implying that the metric does not admit a good boundary description. \\
There is a second useful coordinate system. It does not cover the entire spacetime but only a portion of it, called as the Poincar\'e patch. In fact, we need at least two patches in order to obtain the whole atlas of the manifold. The Poincar\'e patch is defined by local coordinates 
\begin{eqnarray}
x^{i}&=&\ell \frac{x^{i}}{z}\,, \nonumber \\
x^{d}&=&\frac{\ell}{2} \frac{1-x^{2}-z^{2}}{z}\,, \\
x^{d+1}&=& \frac{\ell}{2}\frac{1+x^{2}+z^{2}}{z}\,. \nonumber
\end{eqnarray}
Now, $i=0,\dots,d-1$ and $x^{2}=\eta_{ij}x^{i}x^{j}$. In the embedding, the line element for the Poincar\'e patch becomes
\begin{equation}
   ds^{2}= \frac{\ell ^{2}}{z^{2}}\left(dz^{2}+\eta_{ij}dx^{i}dx^{j}\right) \,. 
\end{equation}
A difference with the previous cases is that the metric is conformally flat and it only covers a triangle on the global patch. Yet, the problem of the divergent behavior at the boundary, now located at $z=0$, persists. Alternatively, we should notice that the metric is finite up to a conformal factor
\begin{equation}
    \hat{g}=z^{2}g \,,
\end{equation}
where $ \hat{g}$ is called \textit{defining} metric which now is smooth at $z=0$. 
The new (conformal) boundary metric is
\begin{equation}
    g_{(0)}=z^{2} g \vert _{z=0} \,.
\end{equation}
In fact, the entire family of spaces that defines a conformal structure at the boundary within a specific topology $\mathbb{R}\times S^{D-2}$ is called asymptotically anti-de Sitter (AAdS) spaces \cite{Ashtekar:1984zz, Henneaux:1985tv}.\\ Let us take a more general metric in Gaussian coordinates (see \autoref{AppendixB}) such that the defining function is in the radial direction $z$
\begin{equation}
    ds^{2}= \frac{1}{z^{2}}\left(dz^{2}+g_{ij}(x,z)dx^{i}dx^{j}\right)\,, 
\end{equation}
where $h_{ij}(x,z)$ is smooth at the boundary $z=0$ by construction. Therefore, it can be expanded in a power series near the boundary as
\begin{equation}
    g_{ij}(x,z)=g_{(0)ij}+z g_{(1)ij}+ z^{2}g_{(2)ij}+\dots \,.
\end{equation}
In General Relativity, odd powers in $z$ are absent, making it easier to work in the Fefferman-Graham frame \cite{fefferman} introducing $\rho=z^{2}$. Then
\begin{equation}
  d{s}^{2}=\frac{{\ell}^{2}}{4{\rho}^{2}}\,{d\rho}^2+{h}_{ij}(x,\rho)d{x}^{i}d{x}^{j} \,. \label{fefferman}
\end{equation}
The induced metric has the pole $1/\rho$, that is, $h_{ij}(x,\rho)=g_{ij}(x,\rho)/\rho$ where $g_{ij}(x,\rho)$  can be expanded in power series around $\rho=0$
\begin{equation}
g_{ij}(x,\rho)=g_{(0)ij}+\rho g_{(2)ij}+\rho^{2}g_{(4)ij}+\cdots \,. \label{fgexpansion}
\end{equation}
Due to the fact that Einstein's equations are second order differential equations, it is necessary to specify the initial data given by $g_{(0)ij}$. Thereby, we can solve the equations iteratively in $\rho$ \cite{deHaro:2000vlm}. However, the initial data it is not enough to determine the coefficient $g_{(d/2)}$ entirely. This near-boundary analysis plays an important role in the renormalization of the AdS action and obtaining holographic quantities as correlation functions and anomaly.

\section{Basic aspects of AdS/CFT correspondence}
As was pointed out above, one of the main interests to pursuit a suitable understanding of gravity theories with AdS asymptotics lies on the Maldacena conjecture.
Concretely, for a supergravity  theory in a low energy regime  with  action $I[\Phi]$ with $\Phi$ as the bulk fields, parametrized by  $\Phi_{(0)}$ on the boundary, the partition function supplemented by the corresponding boundary conditions \cite{Gubser:1998bc} is
\begin{equation}
Z_{SUGRA}[\Phi_{(0)}]=\int_{\Phi \sim \Phi_{(0)}} \mathcal{D}\Phi e^{{-I[\Phi]}}= \left< \exp{\left(-\int\limits_{\partial AAdS}\Phi_{(0)}\mathcal{O}\right)} \right>_{CFT} \,,
\end{equation}
where the right hand side corresponds to the partition function of a CFT on the boundary with operator $\mathcal{O}$ sourced by $\Phi_{(0)}$. Basically, there is a one-to-one correspondence between fields propagating in the bulk on $AdS_{d+1}$ and the operators $\mathcal{O}$ of a CFT defined at the conformal boundary of $AdS$. In the saddle-point approximation, the partition function $Z_{SUGRA} \sim \exp{(-I_{on-shell})}$, thus
\begin{equation}
    I_{on-shell}[\Phi_{(0)}]=-W_{CFT}[\Phi_{(0)}]\,,
\end{equation}
in which $W_{CFT}$ is the quantum effective action of the CFT or the generating function of connected graphs. Knowing the generating function, it is straightforward obtaining the correlation functions of the operator $\mathcal{O}$ determined by functional variations with respect to the source
\begin{equation}
\left< \mathcal{O}(x) \right> =- \frac{\delta W_{CFT} (\Phi_{(0)}) }{\delta \Phi_{(0)}(x)}\bigg\vert _{\Phi_{(0)}=0}\,, 
\end{equation}
or in general an $n$-point function in the AdS/CFT prescription
\begin{eqnarray}
\left< \mathcal{O}(x) \right> &=&  \frac{\delta I_{on-shell}(\Phi_{(0)}) }{\delta \Phi_{(0)}(x)}\bigg\vert_{\Phi_{(0)}=0} \nonumber \\
\left< \mathcal{O}(x_{1}) \mathcal{O}(x_{2})\right> &=& -  \frac{\delta^{2} I_{on-shell}(\Phi_{(0)}) }{\delta \Phi_{(0)}(x_{1}) \delta \Phi_{(0)}(x_{2})}\bigg\vert_{\Phi_{(0)}=0}
\\ 
\left< \mathcal{O}(x_{1})\cdots \mathcal{O}(x_{n})\right> &=& \left(-1\right)^{n+1} \frac{\delta^{n} I_{on-shell}(\Phi_{(0)}) }{\delta \Phi_{(0)}(x_{1})\cdots \delta \Phi_{(0)}(x_{n})}\bigg\vert _{\Phi_{(0)}=0} \nonumber
\end{eqnarray}
Nonetheless, we already said that the metric is large-distance (or IR) divergent, implying also that the gravitational on-shell action is divergent which is seen as UV divergence of the CFT in the duality dictionary. A renormalization scheme which removes these divergences becomes a necessity in order to obtain finite physical quantities.

\section{Action principle and
holographic energy-momentum tensor}
In a manifold with boundary, the Einstein-Hilbert action has to be supplemented with an additional term in order to have a well-defined variational principle.
If we vary the EH action, we find
\begin{equation}
\delta I_{\rm{EH}}= \frac{1}{16\pi G}\int \limits_{\mathcal{M}}^{}d^{d+1}x\sqrt{-g}\mathcal{E}^{\mu}_{\nu}(g^{-1}\delta g)^{\nu}_{\mu} +\frac{1}{16\pi G}\int \limits_{\partial \mathcal{M}}^{}d^{d}x \sqrt{-h} \Theta \,,
\end{equation}
where the last term has been already integrated in the radial coordinate, $h_{ij}$ is the induced boundary metric  and  $\Theta=n_{\mu}\delta^{[\mu \nu]}_{[\alpha \beta]}g^{\beta \rho}\delta\Gamma^{\alpha}_{\nu \rho}$ is a scalar density of the boundary term. Using Gauss-normal coordinates (see \autoref{AppendixB}), $n_{\mu}=(N,\vec{0})$, the on-shell variation becomes
\begin{equation}
  \delta I_{EH}= \frac{1}{16\pi G}\int \limits_{\partial \mathcal{M}}^{}d^{d}x \sqrt{-h}\left( 2\delta K + K^{i}_{j}(h^{-1}\delta h)^{j}_{i}\right)\,.
\end{equation}
The first term corresponds to variation of the extrinsic curvature (\ref{extrinsicK}), i.e, normal derivatives of the metric and the second variations with respect to the boundary metric. Dirichlet boundary conditions corresponds to keeping the induced metric fixed on the boundary, $\delta h=0$. In the General Relativity case, this is not sufficient to properly define the variational principle because of the presence of $\delta K$. Instead,  York and later on Gibbons and Hawking \cite{York:1972sj,Gibbons:1976ue} proposed the addition of a boundary term that trades off $\delta K$ by $\delta h$. In that way, the action that satisfies Dirichlet boundary conditions is
\begin{equation}
    I_{\rm{Dirichlet}}=\frac{1}{16\pi G}\int \limits_{\mathcal{M}}^{}d^{d+1}x\sqrt{-g}\left(R-2\Lambda\right) - \frac{1}{8\pi G}\int \limits_{\partial \mathcal{M}}^{}d^{d}x\sqrt{-h}K\,.
\end{equation}
The variation of the Dirichlet action with respect to the boundary action produces the \textit{quasilocal stress tensor} \cite{Brown:1992br,Creighton:1995au,McGrath:2012db}
\begin{equation}
T^{ij}= \frac{2}{\sqrt{-h}}\frac{\delta I_{Dirichlet}}{\delta h_{ij}}
\end{equation}
which is covariantly conserved at the boundary $\nabla _{i}T^{ij}=0$\footnote{The covariant derivative $\nabla _{i}$ is defined with respect to the induced metric at the boundary $h_{ij}$}. Even so, this definition of energy, in AAdS spacetimes, diverges due to the fact that the induced metric has a pole at the boundary $\rho=0$ reflecting an ill-defined Dirichlet problem for $h_{ij}$. \\
The standard procedure to adress this problem in asymptotically AdS spaces was presented in \cite{Henningson:1998gx,Balasubramanian:1999re,Mann:1999pc,
deHaro:2000vlm} following the same idea of renormalization in quantum field theories. This has been extended to other asymptotic behaviors \cite{Mann:2005yr,Taylor:2008tg}.

If we consider the conformal boundary at certain cutoff $\rho = \epsilon$ in the Fefferman-Graham frame (\ref{fefferman}), we can expand the metric near the horizon $\rho=0$ given by eq.(\ref{fgexpansion}) where the first term $g_{(0)ij}$ corresponds to the initial data in the holographic reconstruction of the spacetime. In the AdS/CFT dictionary, the term $g_{(0)ij}$ is a source of the CFT. Therefore, the functional variation of the action with respect to $g_{(0)ij}$ will give an observable in the dual CFT. Using the source, one can identify the divergent contributions in the action and its variation as combinations of the coefficients $g_{(k)ij}$ of the expansion. Nevertheless, these coefficients are not covariant from the point of view of the bulk manifold, and it is necessary to invert the series in terms of covariant functionals of $g_{ij}(x,\rho)$. In that way, we can write a series of \textit{intrinsic} local counterterms as a boundary term. Added on top of the Dirichlet action, it produces a \textit{finite} action
\begin{equation}
I_{\rm{ren}}=I_{\rm{Dirichlet}}+ \int\limits_{\partial \mathcal{M}}{{d}^{d}}x\sqrt{-h}{\mathcal{L}}_{\mathrm{ct}}(h,\mathcal{R},\nabla \mathcal{R})\,,
\end{equation}
where $\mathcal{R}^{ij}_{lk}(h)$ is the intrisic curvature of the boundary. Now, we can obtain the one-point function, as was anticipated, taking variations with respect to $g_{(0)ij}$ for which there is a well-defined Dirichlet principle
\begin{equation}
\left< T^{ ij }( g_{ (0) } )\right> =\frac { 2 }{ \sqrt { -g_{ (0)ij } }  } \frac { \delta I_{ \mathrm{ ren } } }{ \delta g_{ (0)ij } } =\lim _{ \rho \rightarrow 0 }{ \left( \frac { 1 }{ {\rho }^{ d/2-1 } } { T }^{ ij } [h] \right)  }\,.
\end{equation}
This last quantity is known as the \textit{holographic} energy-momentum tensor \cite{Balasubramanian:1999re}. The energy-momentum tensor contains important holographic information that characterizes the dual CFT. In particular, the conformal anomaly that appears in even dimensional CFTs (odd dimensional bulk) can be computed by tracing the stress tensor.\\
The extensions of the holographic renormalization procedure to gravity theories with higher-order curvature contributions as Quadratic Curvature Gravity or for instance, Lovelock-AdS gravity, are particularly involved and there is no close form for the complete series of counterterms. Instead, we are going to use an alternative scheme, named as Kounterterm method, to renormalized the Lovelock AdS action and obtain finite conserved quantities.

\chapter{Lovelock-AdS gravity} 

\label{Lovelock} 
Despite the success in the predictions of General Relativity since its formulation in 1915, there are still unclear aspects of the spacetime and many modified theories of gravity have been proposed with the intention to achieve a better description of our Universe. \\
In the gauge/gravity picture, as it was shown in the previous chapter, theories with second-order field equation require a minimal set of holographic data.\\
In this way, one of the most natural modifications that preserves this property was proposed by Cornelious Lanczos \cite{Lanczos} in 1938. He considered a precise combination of quadratic terms in the curvature whose variation gives second-order equation of motions 
\begin{equation}
\mathcal{L}_{\rm{GB}}= \sqrt{-g} \left( { R }^{ 2 }+{ R }_{ \mu \nu \alpha \beta  }{ R }^{ \mu \nu \alpha \beta  }-4{ R }_{ \mu \nu  }{ R }^{ \mu \nu  } \right) \,.
\end{equation}
This last Lagrangian, known as Gauss-Bonnet (GB) term, is dynamical in $D>4$. In fact, for $D=4$, the GB term is topological and it does not produce equations of motion. \\
Many years later, David Lovelock \cite{Lovelock:1971yv} generalized the idea of Lanczos, considering a series of dimensionally continued Euler densities 
\begin{equation}
\mathcal{L}_p =\frac{1}{2^p}\,\delta _{\nu 1\cdots \nu _{2p}}^{\mu_1\cdots \mu _{2p}}\,R_{\mu _{1}\mu _{2}}^{\nu _{1}\nu _{2}}\cdots R_{\mu_{2p-1}\mu _{2p}}^{\nu _{2p-1}\nu _{2p}}\,,
\end{equation}
where $\delta _{\nu 1\cdots \nu _{2p}}^{\mu_1\cdots \mu _{2p}}$ is the generalized Kronecker delta of order $p$ (see \autoref{AppendixA}).
 The densities that contribute to the dynamics exists only up to $p=N=[(D-1)/2]$ curvature powers and the entire action can be written as an arbitrary linear combination of them
\begin{equation}
I = \frac { 1 }{ 16\pi G } \int\limits_{ \mathcal{M} }{ d^{ D }x\sqrt { -g } \sum _{ p=0 }^{ N } \alpha _{ p }\mathcal{L}_p\,. }
\end{equation}
In even dimensions, the last term of the series does not contribute to the dynamics of the theory, i.e, it is an Euler topological invariant, similarly to the GB term ($p=2$) in $D=4$. \\
The equations of motion read as
\begin{equation}
\mathcal{E}^{\mu}_{\nu}=\sum _{ p=0 }^{ N } \frac { \alpha _{ p } }{ 2^{ p+1 } }\, \delta _{\nu \nu _{ 1 }\cdots \nu _{ 2p } }^{ \mu \mu _{ 1 }\cdots \mu _{ 2p } }\, R_{ \mu _{ 1 }\mu _{ 2 } }^{ \nu _{ 1 }\nu _{ 2 } }\cdots R_{ \mu _{ 2p-1 }\mu _{ 2p } }^{ \nu _{ 2n-1 }\nu _{ 2p }}=0\,,
\end{equation}
where $\alpha_{p}$ are, in principle, arbitrary coupling constants. There are different criteria to choose them, and one of them would be to maximize or constraint the degrees of freedom of the theory \cite{Teitelboim}. For a given dimension, a generic Lovelock gravity has the same number of degrees of freedom as General Relativity in the same dimension \cite{Dadhich:2015ivt}. In particular cases, however, a number of degrees of freedom can change \cite{Miskovic:2005di} as a consequence of additional symmetries \cite{Banados:1996yj,Giribet:2014hpa}.

Besides, they play an important role in avoiding pathologies such as presence of ghosts, instabilities and causality issues in the AdS/CFT framework \cite{PhysRevLett.55.2656,Konoplya:2017ymp,Konoplya:2017lhs,Camanho:2009hu}.
\\
The theory can be seen as a generalization to Einstein gravity, to include higher-curvature corrections if we identify the first two couplings as $\alpha_{0}=-2\Lambda$ and $\alpha_{1}=0$ that yields
\begin{equation}
I=\frac { 1 }{ 16\pi G } \int  d^{ D }x\sqrt { -g } \left( R-2\Lambda +\sum _{ p=2 }^{ N }{ \frac {\alpha_p }{ 2^{ p } } \, \delta _{ \nu _{ 1 }\cdots \nu _{ 2p } }^{ \mu _{ 1 }\cdots \mu _{ 2p } }\, R_{ \mu _{ 1 }\mu _{ 2 } }^{ \nu _{ 1 }\nu _{ 2 } }\cdots R_{ \mu _{ 2p-1 }\mu _{ 2p } }^{ \nu _{ 2p-1 }\nu _{ 2p } } } \right)\,,
\end{equation}
and whose variation in $D>4$ leads to
\begin{eqnarray}
\mathcal{E}^{\mu}_{\nu}=R_{ \nu  }^{ \mu  }-\frac { 1 }{ 2 } R{ \delta  }_{ \nu  }^{ \mu  }+\Lambda \delta _{ \nu  }^{ \mu  }-H_{ \nu  }^{ \mu  }=0\,.\label{eom}
\end{eqnarray}
Here $H^{\mu}_{\nu}$ is the generalized Lanczos tensor 
\begin{equation}
   H^{\mu}_{\nu}= \sum _{ p=2 }^{ N } \frac { \alpha _{ p } }{ 2^{ p+1 } }\, \delta _{\nu \nu _{ 1 }\cdots \nu _{ 2p } }^{ \mu \mu _{ 1 }\cdots \mu _{ 2p } }\, R_{ \mu _{ 1 }\mu _{ 2 } }^{ \nu _{ 1 }\nu _{ 2 } }\cdots R_{ \mu _{ 2p-1 }\mu _{ 2p } }^{ \nu _{ 2n-1 }\nu _{ 2p }}\,,
\end{equation}
as it includes all higher-curvature corrections to the Einstein tensor with cosmological constant.
Nevertheless, the parameters are completely arbitrary and different families of theories can arise by a proper of choice them. At least, we can distinguish  two more cases of interest: Pure Lovelock (PL) theories \cite{Cai:2006pq,Dadhich:2012ma} and  Lovelock Unique Vacuum (LUV) \cite{Crisostomo:2000bb,Kastor:2006vw}. \\
On the first place, PL theories are obtained turning off the Einstein-Hilbert term, $\alpha_{1}=0$, and leaving only the cosmological constant, $\alpha_{0}\neq 0$ and a particular Lovelock term with $p$ curvature, $\alpha_{p}\neq 0$.
On the other hand, LUV theories are defined by requiring that there is only one and unique constant curvature solution in the theory, implying that all the coupling constants $\alpha$ are functions of only one parameter, $\alpha$ which plays the role of the strength of gravitational interaction.
\\
We are going to discuss properties of these theories in the coming sections.

\section{Constant-curvature solutions and vacuum degeneracy}
Another interesting feature of Lovelock theories of gravity is the rich structure of maximally symmetric solutions. The higher-order curvature terms, in general, modify the characteristic radius $\ell$ of the bare vacuum solution of EH. Instead, a particular vacuum is defined in terms of an effective radius $\ell_{\rm{eff}}$. For Lovelock gravity with negative cosmological constant, the equation of motion can be cast in the form
\begin{equation}
\delta _{ \mu \mu _{ 1 }\cdots \mu _{ 2p } }^{ \nu \nu _{ 1 }\cdots \nu _{ 2p } }\, \left(R_{\nu _{ 1 }\nu _{ 2 } }^{ \mu _{ 1 }\mu _{ 2 } }+\frac{1}{\ell_{\rm{eff}}^{(1) 2 }} \delta_{\nu _{ 1 }\nu _{ 2 } }^{ \mu _{ 1 }\mu _{ 2 } } \right)\cdots  \left(R_{ \nu _{ 2p-1 }\nu _{ 2p } }^{ \mu _{ 2n-1 }\mu _{ 2p }}+\frac{1}{\ell_{\rm{eff}}^{(N) 2}}\delta_{ \nu _{ 2p-1 }\nu _{ 2p } }^{ \mu _{ 2n-1 }\mu _{ 2p }} \right)\,=0\,,
\end{equation}
where $\ell_{\rm{eff}}^{(i)}$, with $i=1,\cdots,N$, are effective AdS radii which are not necessarily different. A particular \textit{global} AdS space within a particular \textit{branch}, with effective radius $\ell_{\rm{eff}}$, is given by
\begin{equation}
 R_{\alpha\beta}^{\mu\nu}=-\frac{1}{\ell_{\rm{eff}}^2}\,\delta_{\alpha \beta}^{\mu \nu}\,.
\end{equation} 
If we insert this condition into the equation of motion (\ref{eom}), we will obtain a polynomial of order $N$ in $\ell_{\rm{eff}}^{-2}$,
\begin{equation}
\Delta (\ell _{\mathrm{eff}}^{-2})=\sum_{p=0}^{N}{\frac{(D-3)!\,(-1)^{p-1}{\alpha }_{p}}{\left( D-2p-1\right) !}}\left( \frac{1}{\ell _{\mathrm{eff}}^{2}}\right) ^{p}=0\,, \label{Delta}
\end{equation}
 such that each root of $\Delta$ defines particular sectors of the theory in the space of parameters $\alpha_{p}$. 
A polynomial of order $N$ with $N$ different solutions defines the existence of non-degenerate theories.
Therefore, it exists an entire set of parameters $\alpha_{p}$ which possess vacua with multiplicity one. In terms of the effective AdS radius, it is equivalent to the condition  
\begin{equation}
\Delta ^{\prime }(\ell _{\mathrm{eff}}^{-2})=\sum_{p=1}^{N}\frac{(D-3)!\,(-1)^{p-1}p\alpha_{p}}{(D-2p-1)!}
\left(\frac{1}{\ell_{\mathrm{eff}}^{2}}\right)^{p-1}\neq 0\,,
\label{DeltaPrime}
\end{equation}
what we will refer to as degeneracy condition of order $1$. We can further understand this condition  with a simple example. Let us consider the case of Einstein-Gauss-Bonnet AdS gravity in $D\ge 5$. We set $p=2$ and $\alpha_{2}=\alpha$ in $H_\nu^\mu$. The polynomial (\ref{Delta}) reads
\begin{equation}
\alpha (D-3)(D-4)\,\frac{1}{\ell_{\rm{eff}}^4}-\frac{1}{\ell_{\rm{eff}}^2} +\frac {1}{\ell^2} =0\,, \label{GB}
\end{equation}
leading to the effective AdS radii
\begin{equation}
\ell^{(\pm)\,2}_{\rm{eff}}=\frac{2\alpha(D-3)(D-4)}{1\pm\sqrt{1-\frac{4 \alpha(D-3)(D-4)}{\ell ^2}}}\,.
\end{equation}
EGB AdS gravity, therefore, has two \emph{simple}, different AdS vacua in the range of Gauss-Bonnet coupling $\alpha<\frac{\ell^2}{4(D-3)(D-4)}$, that is,
\begin{equation}
\ell_{\rm{eff}}^{(+)\,2}\neq \ell_{\rm{eff}}^{(-)\,2}\,. \label{nondegenerate}
\end{equation}
When $\alpha>\frac{\ell^2}{4(D-3)(D-4)}$, the AdS branch does not show up. The theory presents a \textit{critical} point at $\alpha=\frac{\ell^2}{4(D-3)(D-4)}$ where there is only one \emph{degenerate} vacuum of multiplicity two with the radius
\begin{equation}
\ell_{\rm{eff}}^{(+)\,2}=\ell_{\rm{eff}}^{(-)\,2}=\frac{\ell^2}{2}\,,  \label{degenerate}
\end{equation}
and that satisfies $\Delta^{\prime}=0$. In five dimensions, the EGB gravity at the degenerate point in the space of parameters becomes Chern-Simons gravity, which has an enhanced gauge symmetry from Lorentz to AdS. \\
Clearly, for Lovelock-AdS theory, $\Delta^{\prime}=0$ imposes an obstruction to the linearization of the equation of motion. We can prove this by considering a small metric perturbation around a given global AdS branch. The presence of the perturbation is reflected on the curvature 
\begin{equation}
 R_{\alpha\beta}^{\mu\nu}=-\frac{1}{\ell_{\rm{eff}}^2}\,\delta_{\alpha \beta}^{\mu \nu}+\delta \hat{R}_{\alpha \beta}^{\mu \nu} \,,
\end{equation}
where $\delta \hat{R}$ is the contribution coming from the perturbation.
Inserting this into the equation of motion, at the linear order one obtain
\begin{equation}
\mathcal{E}^{\mu}_{\nu}=\Delta^{\prime}(\ell_{\rm{eff}}^{-2})\delta_{\nu \nu_{1}\nu_{2}} ^{\mu \mu_{1}\mu_{2}}(\delta\hat{R})_{\mu_{1}\mu_{2}}^{\nu_{1}\nu_{2}} +\mathcal{O}((\delta\hat{R})^{2}) \,. \label{linearization}
\end{equation}
From here, it is clear that the linear order vanishes when $\Delta^{\prime}=0$. However, it is not the only information that one can extract from this result. The impossibility to define the linear order of the expansion also reflects the obstruction in the definition of a propagating graviton \cite{Camanho:2009hu,Fan:2016zfs} and computation of the black hole mass with perturbative methods \cite{Deser:2002rt,Petrov:2019roe,Kastikainen:2019iaz}. Furthermore, the sign of $\Delta^{\prime}$ determine the sign of the kinetic term in the wave equation \cite{PhysRevLett.55.2656}. Holographic arguments indicate that the requirement $\Delta^{\prime}>0$, which classically implies absence of ghost (or negative energy modes), is analogous to demand unitarity on the CFT side \cite{Camanho:2010ru}.\\
The existence of theories where vacua have higher multiplicity ($\Delta^{\prime}=0$) suggest the possibility of generalizing the result. Indeed, a theory with a vacuum with multiplicity $k$  satisfies the condition that all the derivatives of $\Delta$ vanishes up to order $k$ being the $k$-th one different than zero. Namely,
\begin{eqnarray}
\Delta^{(q)}(\ell_{\mathrm{eff}}^{-2})&=&{\frac{1}{q!}\frac{d^q\Delta}{d(\ell_{\mathrm{eff}}^{-2})^q}}=0\,,\qquad \mathrm{for} \quad  q=1,...,k-1\,, \nonumber \\
\Delta^{(k)}(\ell _{\mathrm{eff}}^{-2})&=&
\sum_{p=k}^{N}\binom{p}{k}\frac{(D-3)!\,(-1)^{p-1}\alpha_{p}\,(\ell _{\mathrm{eff}}^{-2})^{p-k}}{(D-2p-1)!}\neq 0\,.
\label{DeltaK}
\end{eqnarray}
Theories with unique AdS radius (LUV) $\ell=\ell_{\rm{eff}}$ have the maximal degeneracy when $k=N$. In even dimensions $D=2n$ ($k=n-1$), corresponds to Born-Infeld-AdS gravity. The coupling constants have a particular form
\begin{equation}
\alpha_{p}=\binom{n-1}{p}\frac{2^{n-2}(2n-2p-1)!}{\ell^{2(n-p-1)}} \quad \,,0\leq p\leq n-1 \,.
\end{equation}
In odd dimensions $D=2n+1$ ($k=n$), the maximal degenerate case is Chern-Simons-AdS with 
\begin{equation}
\alpha_{p}=\binom{n}{p}\frac{2^{n-1}(2n-2p)!}{\ell^{2(n-p)}}\binom{n}{p} \quad \,,0\leq p\leq n \, \,.
\end{equation}
Even though these two theories seems to have similar properties, their asymptotic behavior, local symmetries and the number of degrees of freedom differ. \\
In the next section, we study the falloff of static solutions in Lovelock-AdS theory with degeneracy $k$.

\section{Static Black Holes}
Spherically symmetric black holes are one of the most simple solutions to any gravity theory. They are described by the following ansatz

\begin{equation}
{\ d }s^{ 2 }=-f(r){\ dt }^{ 2 }+\frac { 1 }{ f(r) } {\ dr }^{ 2 }+{\ r }^{2} \sigma_{mn}(y)dy^mdy^n\,,  \label{ansatz}
\end{equation}
where $\sigma_{nm}$ ($m,n=1,\cdots, D-2$) is the metric of the transversal section of constant curvature $\kappa=+1,0,-1$ . In Lovelock gravity, the metric function $f(r)$ is given an algebraic master equation \cite{PhysRevLett.55.2656,Cai:2001dz,Cai:2003kt} that is the first integral of the gravitational equations of motion, 
\begin{equation}
\sum _{p=0}^{N}\,\frac{\alpha_p(D-3)!}{(D-2p-1)!}\left(\frac{\kappa-f(r)}{r^2}\right)^p=\frac{\mu}{r^{D-1}}\,,
\label{master}
\end{equation}
and which $\mu$ is the integration constant related to the mass of the black hole. For global AdS spaces, we have 
\begin{equation}
    f(r)_{AdS}=\kappa + \frac{r^{2}}{\ell_{\rm{eff}}^{2}}\,. \label{gAdS}
\end{equation}
In asymptotically AdS spaces, $f(r)$ has to behave as
\begin{equation}
f(r)=\kappa+\frac{r^{2}}{\ell_{\rm{eff}}^{2}}+\epsilon (r)\,,
\end{equation}
with $\epsilon (r)$ a function with sufficiently fast falloff, such that global AdS space is recovered (\ref{gAdS}) for large distances. Inserting this into the master equation and Taylor expanding in $1/r$, we find
\begin{eqnarray}
\frac{\mu }{r^{D-1}} &=&\sum_{p=0}^{N}\frac{\left( -1\right) ^{p}(D-3)!\alpha _{p}%
}{\left( D-2p-1\right) !}\,\left( \frac{1}{\ell_{\rm{eff}}^{2}} +\frac{\epsilon }{r^{2}}\right)
^{p}  \nonumber \\
&=&\sum_{q=0}^{N}\sum_{p=q}^{N}\frac{\left( -1\right) ^{p}(D-3)!\alpha _{p}}{%
\left( D-2p-1\right) !}\binom{p}{q}\left(\frac{1}{\ell_{\rm{eff}}^{2}}\right) ^{p-q}\left( \frac{\epsilon }{%
r^{2}}\right) ^{q}\,,
\end{eqnarray}%
or equivalently
\begin{equation}
\frac{\mu }{r^{D-1}}=\sum_{q=0}^{N}\Delta ^{(q)}\,\left( \frac{\epsilon }{%
r^{2}}\right) ^{q}\,,
\end{equation}%
where $\Delta^{(q)}$ is given by eq.(\ref{DeltaK}).  Furthermore, we can expand the series knowing that, for a $k$-degenerate vacuum, the first non-vanishing term is $\Delta ^{(k)} $. Then
\begin{equation}
\frac{\mu }{r^{D-1}}=\Delta ^{(k)}\,\left( \frac{\epsilon }{r^{2}}\right)
^{k}+\Delta ^{(k+1)}\,\left( \frac{\epsilon }{r^{2}}\right) ^{k+1}+\cdots
\,. \label{Expansion1}
\end{equation}%
 It can also be shown that, if $\Delta^{(k+1)}=0$, then all $\Delta^{(q)}=0$  for $q>k$. Having this expansion, we can assume an explicit asymptotic behavior for $\epsilon(r)$ of the form, we get
\begin{equation}
\epsilon (r)=\frac{A}{r^{x}}+\frac{B}{r^{y}}+\cdots \,,\qquad y>x>0\,,\qquad
A\neq 0\,,
\end{equation}%
where $A$, $B$, $x$ and $y$ are coefficients to be determined. Replacing this last equation into (\ref{Expansion1}) and expanding as a binomial form
\begin{equation}
\frac{\mu }{r^{D-1}}=\Delta ^{(k)}\,\left( \frac{A^{k}}{r^{\left( x+2\right)
k}}+\frac{kA^{k-1}B}{r^{\left( x+2\right) \left( k-1\right) +y+2}}+\cdots
\right) +\Delta ^{(k+1)}\,\left( \frac{A^{k+1}}{r^{\left( x+2\right) \left(
k+1\right) }}+\cdots \right) .
\end{equation}%
Clearly, from the leading order, we find
\begin{equation}
\frac{\mu }{r^{D-1}}=\Delta ^{(k)}\,\frac{A^{k}}{r^{\left( x+2\right) k}}\,,
\end{equation}%
and by a simply comparison
\begin{equation}
x=\frac{D-2k-1}{k}\,,\qquad A=\left( -\frac{\mu }{\Delta ^{(k)}}\right) ^{%
\frac{1}{k}}\,. \label{x,A}
\end{equation}%
Notice that the mass parameter may be negative. For the subleading contributions, one obtains
\begin{equation}
0=\Delta ^{(k)}\,\left( \frac{kA^{k-1}B}{r^{\left( x+2\right) \left(
k-1\right) +y+2}}+\cdots \right) +\Delta ^{(k+1)}\,\left( \frac{A^{k+1}}{%
r^{\left( x+2\right) \left( k+1\right) }}+\cdots \right) \,.
\label{sub-leading}
\end{equation}%
There are three possible cases for the coefficients $A,B$:

\emph{a}) When $A=B=0$, the solution becomes the vacuum state of the theory $\mu =0$, i.e, we recover the global AdS space.

\emph{b}) When $A\neq 0$ and $B=0$, the solution exists only if $\Delta
^{(k+1)}=0$. This means that all levels of degeneracy until $\Delta ^{(k)}$, are identically zero. This is the case of Lovelock Unique Vacuum (LUV) theories \cite{Kastor:2006vw, Crisostomo:2000bb}.

\emph{c}) When $A,B\neq 0$, the only way to have non-trivial
coefficients is that both terms (along $\Delta ^{(k)}$ and $\Delta ^{(k+1)}$) in eq.(%
\ref{sub-leading}) are of the same order %
\begin{equation}
y=\frac{2 \left( D-1\right) -2k}{k}\,,  \label{y}
\end{equation}%
and in which case the coefficient is%
\begin{equation}
B=-\frac{\Delta ^{(k+1)}A^{2}}{k\Delta ^{(k)}}=-\frac{\Delta ^{(k+1)}}{%
k\Delta ^{(k)}}\left( \frac{\mu }{\Delta ^{(k)}}\right) ^{\frac{2}{k}}.
\label{B}
\end{equation}%
In this way, for a degenerate vacuum with multiplicity $k$, we find the behavior of the solution as
\begin{equation}
f(r)=\kappa+\frac{r^{2}}{\ell _{\mathrm{eff}}^{2}}+\left( \frac{-\mu }{\Delta ^{(k)}r^{D-2k-1}}\right) ^{\frac{1}{k}%
}-\frac{\Delta ^{(k+1)}}{k {\Delta ^{(k)}}^{2}}\left( \frac{-\mu^{2} }{\Delta ^{(k)}r^{2 \left( D-1\right) -2k}}%
\right) ^{\frac{1}{k}}+\cdots \,. \label{f(r) K} 
\end{equation}%
The formula reproduces known results in particular cases. For example, standard Schwarzschild-AdS solution (EH-AdS) is recovered for $k=1$ and $\Delta^{(k+1)}=0$ 
\begin{equation}
    f_{\rm{Schw-AdS}}=\kappa+\frac{r^{2}}{\ell _{\mathrm{eff}}^{2}}- \frac{\mu }{r^{D-3}}\,.
\end{equation}
The case of LUV theories ($\Delta^{(k+1)}=0$) is also covered. For Born-Infeld AdS $D=2n$ and $k=n-1$, the solution becomes
\begin{equation}
     f_{\rm{BI-AdS}}=\kappa+\frac{r^{2}}{\ell _{\mathrm{eff}}^{2}}+ \left(\frac{-\mu }{r}\right)^{\frac{1}{n-1}} \,.
\end{equation}
The case of Chern-Simons AdS in $D=2n+1$ with $k=n$, is special because the sub-leading terms starting with $r^{(D-2k-1)/k}$ in the metric (\ref{f(r) K}), which vanish in the asymptotic region for other LUV theories, become constant for CS theories when we $k=n$. Thus, the falloff of the metric function drastically changes in CS gravity and it has the slowest falloff from all LUV theories, 
\begin{equation}
    f_{\rm{CS-AdS}}=\kappa+\frac{r^{2}}{\ell_{\mathrm{eff}}^{2}}-
    \mu ^{1/n}\,. \label{CS}
\end{equation}
Both cases were previously reported and extensively studied in the literature \cite{Crisostomo:2000bb,Aros:2000ij,Kastor:2006vw}.
The general form of the asymptotic solution give us the chance to evaluate relevant quantities where physical properties of the theory are codified.

\section{On-shell Weyl tensor vs AdS curvature}
By construction, the Weyl tensor corresponds to the traceless part of the Riemann tensor
\begin{equation}
W_{\alpha \beta}^{ \mu \nu }=R_{ \alpha \beta }^{ \mu \nu }-
\frac { 1 }{ D-2 } \,\delta_{ [\alpha }^{ [\mu }R_{ \beta ] }^{ \nu] }
+\frac {R}{(D-1)(D-2) } \,{\delta }_{ \alpha \beta }^{\mu \nu }\,,
\end{equation}
where
\begin{equation}
 \delta_{[\alpha}^{[\mu}R_{\beta]}^{\nu]}=\delta_{\alpha}^{\mu}R_{\beta}^{\nu}-\delta_{\beta}^{\mu}R_{\alpha}^{\nu}-
\delta_{\alpha}^{\nu}R_{\beta}^{\mu}+\delta_{\beta}^{\nu}R_{\alpha}^{\mu}.
\end{equation}
In standard Einstein gravity, it contains information about the propagating gravitational waves in vacuum. For $D<4$, it vanishes identically.\\
We can compute the on-shell Weyl tensor for EH-AdS by taking the trace of the eq.(\ref{einstein}) (with $T^{\mu}_{\nu}=0$) from where we can obtain the Ricci scalar
\begin{equation}
    R=-\frac{D(D-1)}{\ell^{2}}\,,
\end{equation}
Inserting it back to the equations of motion, we find the Ricci tensor
\begin{equation}
    R^{\mu}_{\nu}=-\frac{(D-1)\delta^{\mu}_{\nu}}{\ell^{2}}\,.
\end{equation}
Taking these results and insert them into the definition of the Weyl tensor that yields its on-shell form for EH-AdS gravity
\begin{equation}
    W^{\mu \nu}_{(EH)\alpha \beta}=R_{ \alpha \beta }^{ \mu \nu }+\frac{1}{\ell^{2}}\delta_{ \alpha \beta }^{\mu \nu }\,.
\end{equation}
On the other hand, we can define the field strength of the $SO(D-1,2)$ group (isomorphic to $AdS_{D}$ group) when the torsional field vanishes, or just AdS curvature as
\begin{equation}
    F_{ \alpha \beta }^{ \mu \nu } =R_{ \alpha \beta }^{ \mu \nu }+\frac{1}{\ell^{2}}\delta_{ \alpha \beta }^{\mu \nu }\,,
\end{equation}
which vanishes for global AdS. Therefore, in EH-AdS gravity, both the on-shell Weyl tensor and AdS curvature coincide
\begin{equation}
    W^{\mu \nu}_{(EH)\alpha \beta}= F_{ \alpha \beta }^{ \mu \nu }\,. \label{wEH}
\end{equation}
In AAdS spacetimes the Ashtekar-Magnon-Das (AMD) \cite{Ashtekar:1984zz,Ashtekar:1999jx} formula,  gives precisely the energy of a black hole in terms of the electric part of the Weyl tensor (see \autoref{AppendixE}). Using (\ref{wEH}), it was shown in ref.\cite{Jatkar:2014npa} that it is possible to reproduce the AMD formula as a Noether charge associated to a properly renormalized EH-AdS action. 
\\
However, this relation does not hold in higher-curvature theories. For Lovelock-AdS theories, the AdS radius is modified 
\begin{equation}
        F_{ \alpha \beta }^{ \mu \nu } =R_{ \alpha \beta }^{ \mu \nu }+\frac{1}{\ell_{\rm{eff}}^{2}}\delta_{ \alpha \beta }^{\mu \nu }\,. \label{AdScurvature}
\end{equation}
 Using the equations of motion (\ref{eom}), one can compare the on-shell Weyl tensor to the AdS curvature in Lovelock-AdS gravity arriving to the result
 \begin{equation}
    W^{\mu \nu}_{\alpha \beta}= F_{ \alpha \beta }^{ \mu \nu }  + X_{ \alpha \beta }^{ \mu \nu } \,.
 \end{equation}
The tensor $X_{ \alpha \beta }^{ \mu \nu }$ vanishes in EH case, and contains higher-curvature terms in general case,
\begin{equation}
    X_{\alpha\beta}^{\mu\nu}=\left( \frac{1}{\ell^2}- \frac{1}{\ell^2_{\mathrm{eff}}}+\frac{2H}{(D-1)(D-2)}\right)
\delta _{\alpha \beta }^{\mu \nu }-\frac{1}{D-2}\,{{\delta }_{[\alpha }^{[\mu }H_{\beta ]}^{\nu ]}}\,.
\label{X}
\end{equation}
Using the solution (\ref{f(r) K} ) and the nonvanishing components of the Riemann tensor (see \autoref{AppendixC}), we can evaluate all these quantities in order to find their asymptotic form. It is straighforward to show that the transversal components of the AdS curvature behave for large $r$ as
\begin{eqnarray}
& F_{m_{1}m_{2}}^{n_{1}n_{2}} =\left( -\frac{\mu }{\Delta ^{(k)}\,r^{D-1}} \right) ^{\frac{1}{k}}\delta _{m_{1}m_{2}}^{n_{1}n_{2}}
+\mathcal{O}\left(r^{-\frac{2(D-1)}{k}}\right)\,, \nonumber \\
&F_{tr}^{tr} =-\frac{(D-2k-1)(D-k-1)}{2k^2}\left( \frac{-\mu }{\Delta ^{(k)}r^{D-1}}\right)^{\frac{1}{k}}
+\mathcal{O}\left( r^{-\frac{2(D-1)}{k}}\right) \,,   \\
&F_{rm}^{rn} =F_{tm}^{tn}=\frac{D-2k-1}{2k}\left( \frac{-\mu }{\Delta ^{(k)}}\right) ^{\frac{1}{k}}r^{-\frac{D-1}{k}}\,
\delta_m^n+\mathcal{O}\left(r^{-\frac{2(D-1)}{k}}\right) \,. \nonumber  \label{AdSasymp}
\end{eqnarray} 

The tensor $X_{ \alpha \beta }^{ \mu \nu }$ can be written as a difference between the Weyl tensor and the AdS curvature. Evaluating (\ref{X}), we find
\begin{equation}
X_{\alpha \beta }^{\mu \nu }\sim (k-1)\,\hat{A}_k\left( \frac{\mu }{r^{D-1}}\right)^{\frac{1}{k}}+(k-2)\,\hat{B}_k\left( \frac{\mu }{r^{D-1}}\right) ^{\frac{2}{k}}+\cdots\,.
\label{W-F}
\end{equation}
The coefficients $\hat{A}_k$ and $\hat{B}_k$ account for higher-order corrections $\alpha _{p\geq 2}$ and they identically vanish in EH gravity
($\alpha _{p\geq 2}=0$).
The above expression singles out $k=1$ as a special case with particular asymptotic behavior of the Weyl tensor.\\
Consequently, the difference between the Weyl tensor and the AdS curvature in any Lovelock AdS gravity is
subleading in $r$ of order $X_{\alpha \beta }^{\mu \nu }=\mathcal{O}\left(\mu ^{2}/r^{2(D-1)/k}\right) $. This is in agreement with the known result in EGB AdS gravity \cite{Jatkar:2015ffa} for $k=1$.
Then, the leading order behavior of the Weyl tensor is
\begin{equation}
W_{\alpha\beta}^{\mu\nu} \sim F_{\alpha\beta}^{\mu\nu} \sim \mathcal{O}\left( 1/r^{\frac{D-1}{k}}\right) \,. \label{WeylAsymp}
\end{equation}
In the next section, we are going to prove that for a given theory with degeneracy $k$, the conserved charge formula must contain $k$ powers of the Weyl tensor in order to produce a finite result  that can be interpreted as the mass of the corresponding black hole solution. 


\include{Chapters/Kounterterms}

\chapter{Kounterterm Method and Conserved Charges} 

\label{NoetherTheorem} 


The renormalization of gravity in AAdS spaces, as emphasized in \autoref{AdS/CFT}, is essential in the holographic computations. It allows to regulate the UV divergences of the dual CFT by dealing with the IR divergences of the gravity theory. Nevertheless, the standard procedure is based on a perturbative approach and the level of complexity increases considerably when higher-curvature terms are included.\\
An alternative method, called Kounterterm renormalization \cite{Olea:2005gb,Olea:2006vd}, considers the addition of a boundary term whose existence is linked to topological invariants and transgression forms. \\
It is inspired by the fact that the addition of topological invariants to the gravity action modifies the boundary dynamics and not the bulk dynamics. Concrete evidence shows
that the Einstein-Hilbert AdS action coupled to Gauss-Bonnet, in $D=4$, produces correct conserved and thermodynamic quantities \cite{Aros:1999id,Olea:2005gb}. \\
In any dimensions, the renormalized action has the form
\begin{equation}
I_{\mathrm{ren}}={I}_{\mathrm{bulk}}+{c}_{D-1}\int\limits_{\partial {\mathcal{M}}}{{d}^{D-1}x\,{B}_{D-1}}(h,K,\mathcal{R})\,,
\label{Ireg}
\end{equation}
where $B_{D-1}(h,K,{\mathcal{R}})$ is a boundary term that depends on the boundary metric, the extrinsic curvature, the boundary curvature and $c_{D-1}$ is coupling constant fixed by requiring a well-defined variational principle. The boundary term has an universal form regardless the theory under consideration being valid from Einstein-Hilbert AdS to any Lovelock-AdS theory. However, the coupling constant must be fixed depending on the theory \cite{Kofinas:2007ns,Kofinas:2008ub}.\\
Finite holographic quantities such as correlation functions \cite{Anastasiou:2017mag}, entanglement entropy and R\'enyi entropies have been obtained using this prescription \cite{Anastasiou:2018rla,Anastasiou:2018mfk}. Another interesting feature are the conserved quantities. The Kounterterm renormalization produces finite Noether charges for rotating black holes \cite{Olea:2005gb,Diaz-Martinez:2018utg} and, also, higher-derivative gravity theories \cite{Anastasiou:2017rjf,Giribet:2018hck}. \\
In the extension to Lovelock-AdS theories in $D=2n$ dimensions, the Kounterterm series reads
\begin{eqnarray}
B_{2n-1}=2n\sqrt {-h} \int\limits_{0}^{1} du\,
\delta_{j_1\dots j_{ 2n-1 } }^{i_1\dots i_{2n-1}}K_{i_1}^{j_1}
\left( \frac{1}{2} \mathcal{R}_{i_2 i_3}^{j_2 j_3}-u^2  K_{i_2}^{j_2} K_{i_3}^{j_3} \right) \times \dots  \\
\dots \times \left( \frac{1}{2} \mathcal{R}_{i_{2n-2} i_{2n-1}}^{j_{2n-2} j_{2n-1}}-u^2  K_{i_{2n-2}}^{j_{2n-2}} K_{i_{2n-1}}^{j_{2n-1}} \right),  \nonumber
\label{evenk}
\end{eqnarray}
and corresponding coupling constant is
\begin{equation}
c_{2n-1}=-\frac {1}{16\pi nG}\sum _{p=1}^{n-1}
\frac{p\alpha_p}{(D-2p)!}{(-\ell_{\mathrm{eff}}^2)^{n-p}}\,.
\end{equation}

In $D=2n+1$ dimensions, the boundary term takes the form
\begin{eqnarray}
B_{2n}=2n\sqrt{-h}\int\limits_{0}^{1}du\int\limits_{0}^{u}dt\,
\delta_{j_1\dots j_{2n-1}}^{i_1\dots i_{2n-1}} K_{i_1}^{j_1}
\left(\frac{1}{2}\,\mathcal{R}_{i_2 i_3}^{j_2j_3}-u^2K_{i_2}^{j_2}
K_{i_3}^{j_3}+\frac{t^2}{\ell_{\mathrm{eff}}^2}\,\delta_{i_2}^{j_2}
\delta_{i_{ 3 } }^{j_{ 3 } } \right) \times \dots \nonumber \\
\dots \times \left(\frac{1}{2}\,\mathcal{R}_{i_{2n-2}{i}_{2n-1}}^{j_{2n-2}j_{2n-1}}
-u^2 K_{i_{2n-2}}^{j_{2n-2}}
K_{i_{2n-1}}^{j_{2n-1}}+\frac{t^2}{\ell_{\mathrm{eff}}^2}
\delta_{i_{2n-2}}^{j_{2n-2}}\,\delta_{i_{2n-1}}^{j_{2n-1}}\right), 
\end{eqnarray}
with the coupling constant
\begin{equation}
c_{ 2n }=-\frac { 1 }{ 16\pi n G } \left[ \int\limits_{ 0 }^{ 1 } du (1-u^2)^{n-1}  \right]^{ -1 }
\sum _{ p=1 }^{ n }%
\frac{ p\alpha_{ p } }{( D-2p)! } \,(-\ell_{\mathrm{eff}}^2)^{n-p}\,.
\label{c_odd}
\end{equation}

\section{Noether theorem}
The Noether theorem \cite{Noether:1918zz} is one of the building blocks of modern physics. The theorem relates the existence of continuous symmetries and conservation laws in a system. This issue has contributed significantly to the development of many physical theories during the last 100 years.\\
Yet, the mechanism of applying the theorem to General Relativity was unclear until  1993 when R. Wald \cite{Wald:1993nt} gave a prescription for it.  The idea of Wald is based on the invariance of gravity theories under general coordinate transformations ( diffeomorphisms), and the spacetime isometries described by Killing vectors. \\
For a generic Lagrangian density $\mathcal{L}(g,R)$ and a set of Killing vectors $\{\xi^\mu\}$, the Noether current can be constructed by taking diffeomorphic variations $\delta_\xi$ of $\mathcal{L}$ with respect to the fields and derivatives of the fields. These variations corresponds to Lie derivatives along an isometry of the theory that corresponds to a Killing vector. In fact, the variations of the metric with respect to infinitesimal coordinate transformations is essentially the Killing equation
\begin{equation}
    \delta_{\xi} g_{\mu \nu}=-\pounds_{\xi}g_{\mu \nu}=-(\nabla _{\mu}\xi_{\nu}+\nabla_{\nu}\xi_{\mu})\,. \label{metricvary}
\end{equation}
On the other hand, the variations of the Christoffel symbol has the form 
\begin{eqnarray}
\delta _{ \xi  }\, \Gamma _{ \beta \lambda  }^{ \mu  }=-\pounds_{ \xi  }\Gamma _{ \beta \lambda  }^{ \mu  }=-\frac { 1 }{ 2 } \left( \nabla _{ \beta  }\nabla _{ \lambda  }\xi ^{ \mu  }+\nabla _{ \lambda  }\nabla _{ \beta  }\xi ^{ \mu  } \right) - \frac{1}{2} \left(R_{ \lambda \sigma \beta  }^{ \mu  } + R^{\mu}_{\beta \sigma  \lambda}\right)\xi^{\sigma}\,. \label{LieGamma}
\end{eqnarray}
Using the above expressions, we can write the variations of the  renormalized Lovelock-AdS action (\ref{Ireg}) that was carry out by Kofinas et al. \cite{Kofinas:2007ns,Kofinas:2008ub}. This leads to
\begin{equation}
    \delta_{\xi}I_{\rm{ren}} = \delta_{\xi}I_{\rm{bulk}}+ c_{D-1}\int\limits_{\partial {\mathcal{M}}}{{d}^{D-1}x\,\delta_{\xi}{B}_{D-1}}\,,
\end{equation}
where the first term on the right-hand side is given by
\begin{equation}
 \delta_{\xi}I_{\rm{bulk}}=  \int\limits_{ \mathcal{M}}{{d}^{D}x\,\pounds_{\xi}\left(\sqrt{-g}\mathcal{L}_{\rm{bulk}}\right)+\partial_{\mu}\left(\sqrt{-g}\xi^{\mu}\mathcal{L}_{\rm{bulk}}\right)}\,.
\end{equation}
On-shell, we can identify the surface contributions and using a radial foliation of the spacetime $\mathcal{M}$ in Gauss-normal coordinates
\begin{equation}
ds^2=N^2(r)\,dr^2+h_{ij}(r,x)\,dx^idx^j\,,
\end{equation}
where $N(r)$ is the lapse function and $h_{ij}$ is the induced metric at a fixed $r$. The outward-pointing normal to the boundary is $n_{\mu}=(n_{r},n_{i})=(N,\vec{0})$. The variation becomes
 \begin{equation}
 \delta_{\xi}I_{\rm{ren}}=  \int\limits_{ \partial \mathcal{M}}{{d}^{D-1}x\,n_{\mu}\left(\frac{1}{N}\Theta^{\mu}+\sqrt{-h}\xi^{\mu}\mathcal{L}_{\rm{bulk}}+c_{D-1}n^{\mu}\partial_{i}(\xi^{i}B_{D-1})\right)}\,,
\end{equation}
where $\Theta(\xi)=\frac{1}{N}n_{\mu}\Theta^{\mu}$ are the boundary terms associated to the bulk Lagrangian. Therefore, the Noether current can be identified as
\begin{equation}
    \mathcal{J}^{\mu}=\frac{1}{N}\Theta^{\mu}+\sqrt{-h}\xi^{\mu}\mathcal{L}_{\rm{bulk}}+c_{D-1}n^{\mu}\partial_{i}(\xi^{i}B_{D-1})\,. \label{current}
\end{equation}
As the current is covariantly conserved, we can introduce the Noether charge as an integral of the current projected on a normal direction,
\begin{equation}
Q[\xi]=  \int\limits_{ \partial \mathcal{M}}{{d}^{D-1}x\,\frac{1}{N}n_{\mu}\mathcal{J}^{\mu} }
\end{equation}
 In order to obtain the energy of a black hole, the integral must be over a co-dimension 2 surface at fixed time and radial infinity and with a time-like Killing vector.

\section{Conserved Charges in Lovelock-AdS gravity}

The boundary metric $h_{ij}$ admits a time-like ADM foliation 
\begin{equation}
h_{ij}dx^{i}dx^{j}=- \tilde{N}^{2}(t)dt^{2}+\sigma_{mn}\left(dy^{m} +\tilde{N}^{m}dt\right)\left(dy^{n} +\tilde{N}^{n}dt\right)\,,
\end{equation}
where now $\sqrt{-h}=\tilde{N}\sqrt{\sigma}$, with $\sigma_{mn}$ the codimension-2  metric of the asymptotic boundary $\Sigma_{\infty}$ at constant time, and $\sigma$ is the determinant. The
unit normal to the hypersurface is given by $u_{j}=(u_{t},u_{m})=(-\tilde{N},0)$ and the conserved current (\ref{current}), projected to the codimension-2 surface, can be written as a tensorial density $\hat{\tau}^{i}_{j}$. Therefore, the conserved charges are given by the surface integral
\begin{equation}
Q(\xi )=\int\limits_{\Sigma _{\infty }}{{d}^{D-2}x\sqrt{\sigma }\,{u}_{j}\hat{\tau}^{j}_{i}{\xi }^{i}}\,.  \label{Q}
\end{equation}
The charge density is naturally split in two contributions:
\begin{equation}
\hat{\tau}_{i}^{j}  =\tau _{i}^{j}+\tau _{(0)i}^{j}\,. \label{tau}
\end{equation}
Here $\tau _{i}^{j}$, when integrated, can be identified with the mass and angular momentum of the black hole. In the case of non-degenerate theories, it was argued in refs.\cite{Kofinas:2007ns, Arenas-Henriquez:2017xnr} that it is factorizable by the Weyl tensor because it vanishes for global AdS space\footnote{We pointed out in the previous section that Chern-Simons AdS gravity is the only exception in this procedure and they will be discussed later.}. That means, it vanishes for vacuum configurations of the solution space, i.e., global AdS.
In even dimensions $D=2n$ as it is presented in \cite{Kofinas:2007ns,Kofinas:2008ub}, $\tau _{(0)i}^{j}$ is identically zero, and the only contribution comes from
\begin{align}
\tau ^{ j }_{ i } =  & \frac { 1 }{2^{ n-2 } }  \delta ^{ [j{ j }_{ 2 }\dots { j }_{ 2n-1 }] }_{ [{ i }_{ 1 }{ i }_{ 2 }\dots { i }_{ 2n-1 }] }\, K^{ { i }_{ 1 } }_{ i }   \bigg[\frac { 1 }{16\pi G}\sum _{ p=1 }^{ n-1 } \frac { p{ \alpha _{ p } } }{ (2n-2p+1)! }R_{ j_{ 2 }j_{ 3 } }^{ i_{ 2 }i_{ 3 } } \times \cdots  \nonumber \\ &
   \cdots \times R_{ j_{ 2p-2 }j_{ 2p-1 } }^{ i_{ 2p-2 }i_{ 2p-1 } }  \delta _{ j_{ 2p }j_{ 2p+1 } }^{ i_{ 2p }i_{ 2p+1 } }\cdots \delta _{ j_{ 2n-2 }j_{ 2n-1 } }^{ i_{ 2n-2 }i_{ 2n-1 } }+ 
 n{ c }_{ 2n-1 }R_{ j_{ 2 }j_{ 3 } }^{ i_{ 2 }i_{ 3 } }\cdots R_{ j_{ 2n-2 }j_{ 2n-1 } }^{ i_{ 2n-2 }i_{ 2n-1 } } \bigg]\,.
\label{qqeven}
\end{align}
In contrast, in odd dimensions $D=2n+1$ both contributions of (\ref{tau}) are present. They are given by
\begin{align}
\tau ^{ j }_{ i }= & \frac { 1 }{ 2^{ n-2 } }  \delta ^{ [j{ j }_{ 2 }\dots { j }_{ 2n }] }_{ [{ i }_{ 1 }{ i }_{ 2 }\dots { i }_{ 2n }] }\, K^{ { i }_{ 1 } }_{ i }   { \delta  }_{ { j }_{ 2 } }^{ { i }_{ 2 } }\bigg[ \frac { 1 }{ 16\pi G } \sum _{ p=1 }^{ n }{ \frac { p{ \alpha _{ p } } }{ (2n-2p)! } R_{ j_{ 3 }j_{ 4 } }^{ i_{ 3 }i_{ 4 } }\cdots R_{ j_{ 2p-1 }j_{ 2p } }^{ i_{ 2p-1 }i_{ 2p } } } \delta _{ j_{ 2p+1 }j_{ 2p+2 } }^{ i_{ 2p+1 }i_{ 2p+2 } }\cdots \delta _{ j_{ 2n-1 }j_{ 2n } }^{ i_{ 2n-1 }i_{ 2n } }+  \nonumber \\&
+n{ c }_{ 2n }\int\limits_{0}^{1} {dt \left( R_{ j_{ 3 }j_{ 4 } }^{ i_{ 3 }i_{ 4 } }+\frac { { t }^{ 2 } }{  \ell_{\rm{eff}}^{ 2 } } \delta _{ j_{ 3 }j_{ 4 } }^{ i_{ 3 }i_{ 4 } } \right) \cdots  } \left( R_{ j_{ 2n-1 }j_{ 2n } }^{ i_{ 2n-1 }i_{ 2n } }+\frac { { t }^{ 2 } }{ \ell_{\rm{eff}}^{ 2 } } \delta _{ j_{ 2n-1 }j_{ 2n } }^{ i_{ 2n-1 }i_{ 2n } } \right)  \bigg]\,, \label{qqodd} 
\end{align}
and 
\begin{eqnarray}
\tau _{ (0)i }^{ j }=-\frac { nc_{ 2n } }{ 2^{ n-2 } } \, \int _{ 0 }^{ 1 } du\, u\, \delta _{ [{ i }_{ 1 }{ i }_{ 2 }...{ i }_{ 2n }] }^{ [j{ j }_{ 2 }...{ j }_{ 2n }] }\left( { \delta  }_{ { j }_{ 2 } }^{ { i }_{ 2 } }K_{ i }^{ { i }_{ 1 } }+{ \delta  }_{ { i } }^{ { i }_{ 2 } }{ K }_{ { j }_{ 2 } }^{ { i }_{ 1 } } \right) { \mathcal{F} }_{ { j }_{ 3 }{ j }_{ 4 } }^{ { i }_{ 3 }{ i }_{ 4 } }\times \cdots \times { \mathcal{F }}_{ { j }_{ 2n-1 }{ j }_{ 2n } }^{ { i }_{ 2n-1 }{ i }_{ 2n } }\,,
\end{eqnarray}
where $\mathcal{F}_{ kl}^{ij}$ is a function of the parameter $u$
\begin{equation}
\mathcal{F}_{ kl}^{ij}={ R }_{ kl}^{ ij }-\left( u^{ 2 }-1 \right)\left( K^{i}_{k}K^{j}_{l}-K^{i}_{l}K^{j}_{k}   \right)+\frac { u^{ 2 } }{ \ell _{ \rm{ eff } }^{ 2 } }  { \delta  }_{{[kl]}}^{{[ij]}  }
\end{equation}
   Notice that  $\tau_{(0)i}^{j}$ is not factorizable by the AdS curvature. Thus, $\tau_{ (0)i }^{ j }$ gives rise to the vacuum energy of the corresponding system.  
  The energy of global AdS space is shifted respect to the Hamiltonian mass by the presence of a vacuum energy. This vacuum energy can be related to the Casimir energy of the dual CFT using the standard dictionary of the couplings given by gauge/gravity correspondence.\\
    A power-counting argument suggests the asymptotic falloff of $\tau_{i}^{j}$ necessary in order to produce a finite energy for the theory. 
    If the bulk metric asymptotically behaves given by eq.(\ref{f(r) K}), then we have that
    $\sqrt{\sigma}\sim \mathcal{O}(r^{D-2})$, $u_{j}\sim \mathcal{O}(r)$  and $\xi\sim \mathcal{O}(1)$. The full charge $Q(\xi)$ must be of order $\mathcal{O}(1)$, thus $\tau _{i}^{j}$ should be of order $\mathcal{O}(1/r^{D-1})$ to produce a finite quantity.
      In what follows, we manipulate the expression for $\tau_{i}^{j}$ in order to make manifest the dependence on the degeneracy conditions which appear at different orders.\\

\subsection{Even dimensions $D=2n$}
The charge density in even dimensions can be expressed as
\begin{eqnarray}
\tau_{i}^{j} &=&\frac{{{\ell _{\mathrm{eff}}^{2n-2}}}}{16\pi G{2}%
^{n-2}}\,\delta
_{i_{1}\cdots i_{2n-1}}^{jj_{2}\cdots
j_{2n-1}}\,K_{i}^{i_{1}}\sum_{p=1}^{n-1}{\frac{p{\alpha }_{p}}{\left( 2n-2p\right) !}}\left( \frac{1}{{{\ell _{\mathrm{eff}}^{2}}}}%
\right) ^{p-1}R_{j_{2}j_{3}}^{i_{2}i_{3}}\cdots
R_{j_{2p-2}j_{2p-1}}^{i_{2p-2}i_{2p-1}}\times  \nonumber \\
&& \left[ {\left( \frac{1}{{{\ell _{\mathrm{eff}}^{2}}}}\right)
^{n-p}\delta _{j_{2p}j_{2p+1}}^{i_{2p}i_{2p+1}}}\cdots \delta
_{j_{2n-2}j_{2n-1}}^{i_{2n-2}i_{2n-1}}-{(-1)}%
^{n-p}R_{j_{2p}j_{2p+1}}^{i_{2p}i_{2p+1}}\cdots
R_{j_{2n-2}j_{2n-1}}^{i_{2n-2}i_{2n-1}}\right] \,. \label{echarge}
\end{eqnarray}
where we have used the explicit form of the coupling $c_{2n-1}$. The above expression is in fact a polynomial of order $n-1$ in the Riemann curvature. For a particular vacuum with degeneracy $k$, this may be rearranged in order to express the polynomial as a product of $k$  AdS curvatures times a polynomial  $\mathcal{P}(R)$ of order $n-1-k$ in the curvature
\begin{align}
\tau ^{ j }_{ i }=\frac { { { \ell _{ \rm{ eff } }^{ 2(n-1) } } } }{ 16\pi G\, { 2 }^{ n-2 } } \,\delta _{ i_{1} i_{ 2 }\cdots { i }_{ 2k }{ i }_{ 2k+1 }\dots { i }_{ 2n }i_{ 2n-1 } }^{ jj_{ 2 }\cdots { j }_{ 2k }{ j }_{ 2k+1 }\dots { j }_{ 2n }j_{ 2n-1 } }\, \, K_{ i }^{ i_{ 1 } }\, \left( R_{ j_{ 2 }j_{ 3 } }^{ i_{ 2 }i_{ 3 } }+\frac { 1 }{ \ell _{ \rm{ eff } }^{ 2 } } \delta _{ j_{ 2 }j_{ 3 } }^{ i_{ 2 }i_{ 3 } } \right) \times \dots \nonumber \\
 \dots  \times \left( R_{ j_{ 2k }j_{ 2k+1 } }^{ i_{ 2k }i_{ 2k+1 } }+\frac { 1 }{ \ell _{ \rm{ eff } }^{ 2 } } \delta _{ j_{ 2k }j_{ 2k+1 } }^{ i_{ 2k }i_{ 2k+1 } } \right) { \mathcal{P}_{ j_{ 2k+2 }\cdots j_{ 2n-1 } }^{ i_{ 2k+2 }\cdots i_{ 2n-1 } } }(R), \label{goal}
\end{align}
where $\mathcal{P}(R)$ (see explicit construction in Appendix \ref{evenfactorization}) reads
\begin{eqnarray}
&&\left. \mathcal{P}{_{j_{2k+2}\cdots j_{2n-1}}^{i_{2k+2}\cdots i_{2n-1}}}%
(R)=\sum_{s=0}^{n-k-1}\sum_{p=k}^{k+s}\Delta ^{(p)}(C_{k+s,p}){%
_{j_{2k+2}\cdots j_{2n-2s-1}}^{i_{2k+2}\cdots i_{2n-2s-1}}\times }\right.
\notag \\
&&\hspace{-0.3cm}\left( R_{j_{2n-2s}j_{2n-2s+1}}^{i_{2n-2s}i_{2n-2s+1}}+%
\frac{1}{\ell _{\mathrm{eff}}^{2}}\delta
_{j_{2n-2s}j_{2n-2s+1}}^{i_{2n-2s}i_{2n-2s+1}}\right) \cdots \left(
R_{j_{2n-2}j_{2n-1}}^{i_{2n-2}i_{2n-1}}+\frac{1}{\ell _{\mathrm{eff}}^{2}}%
\delta _{j_{2n-2}j_{2n-1}}^{i_{2n-2}i_{2n-1}}\right)  \label{P}
\end{eqnarray}
In general, the construction of the coefficients $C_{ k+s,p }$ is involved. As it would become clear later, for our purpose, the problem reduces to obtaining a single coefficient. \\
In order to obtain the mass, we must evaluate the charge in the asymptotic region where the bulk curvature tensor behaves as in eq.(\ref{Riem}) and the extrinsic curvature behaves as
\begin{equation}
K_{j}^{i}  \sim  -\frac{1}{\ell _{\mathrm{eff}}}\,\delta _{j}^{i}+\mathcal{O}(1/r^{2})\,. \label{extrinsic}
\end{equation}
As we argued in the last section, the correct behavior of the charge must be $\tau^{i}_{j} \sim \mathcal{O}(1/r^{D-1})$. Given the falloff of the AdS curvature in eq.(\ref{AdSasymp}), the polynomial is reduced to order $\mathcal{P}\sim \mathcal{O}(1)$. This amounts to evaluate all the curvatures in the polynomial just at the leading order. In this case, the only nonvanishing the coefficient is
\begin{equation}
(C_{kk}){_{j_{2k+2}\cdots j_{2n-1}}^{i_{2k+2}\cdots i_{2n-1}}}=\frac{%
(-1)^{k-1}}{2(2n-3)!\ell _{\mathrm{eff}}^{2(n-2)}}\,\delta {%
_{j_{2k+2}j_{2k+3}}^{i_{2k+2}i_{2k+3}}}\cdots \delta {_{j_{2n-2}\cdots
j_{2n-1}}^{i_{2n-2}i_{2n-1}}}
\end{equation}
 what leads to
\begin{equation}
\mathcal{P}_{j_{2k+2}\cdots j_{2n-1}}^{i_{2k+2}\cdots i_{2n-1}}(-\ell_{\mathrm{eff}}^{-2}\delta^{[2]})=-\frac{(-1)^{k}\Delta^{(k)}}{2(2n-3)!\ell_{\rm{eff}}^{2(n-2)}}\,\delta_{{j}_{2k+2}{j}%
_{2k+3}}^{{i}_{2k+2}{i}_{2k+3}}\cdots {\delta }_{{j}_{2n}{j}_{2n-1}}^{{i}%
_{2n}{i}_{2n-1}}.
\end{equation}%
Evidently, this quantity is valid only on an AAdS branch of degeneracy $k$. Replacing the asymptotic form of the extrinsic curvature given by eq.(\ref{extrinsic}) and the relation between the AdS curvature and the Weyl tensor as shown in eq.(\ref{WeylAsymp}), we find that
\begin{eqnarray}
\tau _i^j &=&\frac{\ell_{\mathrm{eff}}(-1)^k\,\Delta ^{(k)}}{16\pi G\,{2}^{n-1}(2n-3)!}\,\delta _{ii_{2}\cdots {i}_{2k}{i}_{2k+1}\dots {i}%
_{2n}i_{2n-1}}^{jj_{2}\cdots {j}_{2k}{j}_{2k+1}\dots {j}_{2n}j_{2n-1}}\,{W}_{%
{j}_{2}{j}_{3}}^{{i}_{2}{i}_{3}}\cdots {W}_{{j}_{2k}{j}_{2k+1}}^{{i}_{2k}{i}%
_{2k+1}}\times  \notag \\
&&\qquad \qquad \qquad \qquad \qquad \times {\delta }_{{j}_{2k+2}{j}%
_{2k+3}}^{{i}_{2k+2}{i}_{2k+3}}\cdots {\delta }_{{j}_{2n}{j}_{2n-1}}^{{i}%
_{2n}{i}_{2n-1}}\,.
\end{eqnarray}%
Contracting the Kronecker deltas, the charge adquires the form of $k$-th power in the Weyl tensor, as expected,
\begin{equation}
\tau _{i}^{j}=\frac{{{\ell_{\mathrm{eff}}{ (-1)}^{k}\,\Delta
^{(k)}(2n-2k-2)!}}}{16\pi G\,{2}^{k}(2n-3)!}\,\delta _{ii_{2}\cdots {i}%
_{2k}{ i}_{2k+1}}^{jj_{2}\cdots {j}_{2k}{ j}_{2k+1}}\,{W}_{{ j}_{2}{
j}_{3}}^{{ i}_{2}{\ i}_{3}}\cdots {W}_{{ j}_{2k}{j}_{2k+1}}^{{ i}_{2k}%
{ i}_{2k+1}}\,.\label{teven}
\end{equation}
Remarkably, for $k=1$, the last equation reduces to a known linear expression in the Weyl tensor. As a matter of fact, the above formula corresponds to the generalization of the Ashtekar-Magnon-Das mass \cite{Ashtekar:1984zz, Ashtekar:1999jx} for nondegenerate Lovelock-AdS gravity (see \autoref{AppendixE}).\\

\subsection{Odd dimensions $D=2n+1$}
We proceed similarly as in even dimensions, considering eq.(\ref{qqodd}) rewritten in a convenient way as
\begin{equation}
\tau _i^j =\frac{(2n-1)!(-\ell _{\mathrm{eff}}^2)^{n-1}}{2^{3n-4}(n-1)!^2
16\pi G}\, \delta _{ i_{ 1 }\cdots i_{ 2n } }^{ jj_{ 2 }\cdots j_{ 2n } }K_{
i }^{ i_{ 1 } }\delta _{ j_{ 2 } }^{ i_{ 2 } }\sum _{ p=1 }^{ n } \frac {
(-1)^{ p }p\alpha _{ p }\, \ell _{ \mathrm{{eff } }}^{ 2(1-p) } }{
(2n-2p+1)! } \int _{ 0 }^{ 1 } du\, (\mathcal{I}_{p})_{ j_{ 3 }\cdots j_{
2n } }^{ i_{ 3 }\cdots i_{ 2n } }(u)\,,   \label{qodd}
\end{equation}
where the tensorial quantity $\mathcal{I}_{p}(u)$ is a polynomial in the curvature
\begin{equation}
\mathcal{I}_{p}(u)=\left( R+\frac{u^{2}}{\ell _{\mathrm{eff}}^{2}}\,\delta
^{\lbrack 2]}\right) ^{n-1}-(u^2-1) ^{n-1}(-R)^{p-1}\left( \frac{1}{\ell _{%
\mathrm{eff}}^{2}}\,\delta ^{\lbrack 2]}\right) ^{n-p}.   \label{I}
\end{equation}
The limiting case corresponds to Chern-Simons AdS gravity which is the maximally degenerate theory in odd dimensions ($k=n$). As previously observed, this case is special and $\tau_{i}^{j}$ vanishes identically. Thus, we are going to restrict our analysis for theories with $k<n$.  We shall rearrange the expression as in the even dimensional case, casting the charge density in the form
\begin{eqnarray}
\tau _i^j&=&\frac{ 2^{2n-2}{(n-1)!}^2\ell _{ \mathrm{eff}}^{ 2(n-1) } }{ 2^{ n-2 }(2n-1)! 16\pi G}\,\delta _{ i_{ 1 }i_{ 2 }\cdots {%
 i }_{ 2k }{ i }_{ 2k+1 }\dots { i }_{ 2n }i_{ 2n-1 } }^{ jj_{ 2 }\cdots {%
 j }_{ 2k }{j }_{ 2k+1 }\dots { j }_{ 2n }j_{ 2n-1 } }K_{ i }^{ i_{ 1 }
}\left( R_{ j_{ 2 }j_{ 3 } }^{ i_{ 2 }i_{ 3 } }+\frac { 1 }{ \ell _{ \mathrm{eff}}^2}\, \delta _{ j_{ 2 }j_{ 3 } }^{ i_{ 2 }i_{ 3 } } \right)
\times \dots  \notag \\
&& \qquad \qquad \dots \times \left( R_{j_{ 2k }j_{ 2k+1 } }^{i_{ 2k }i_{
2k+1 } }+\frac { 1 }{ \ell _{\mathrm{{eff } }}^{ 2 } }\, \delta _{j_{ 2k
}j_{2k+1 } }^{i_{ 2k }i_{ 2k+1 } } \right) \mathcal{ P }_{j_{
2k+2 }\cdots j_{ 2n-1 } }^{ i_{2k+2 }\cdots i_{2n-1 } }(R)\,,
\label{qoddfact}
\end{eqnarray}
where once again, $\mathcal{P}(R)$ is a polynomial in the curvature tensor of order $n-k-1$. Using the same reasoning as in the even dimensional case, it is possible to write the polynomial $\hat{\mathcal{P}}$ in terms of the degeneracy condition functions $Delta^{(p)}$ order by order in $p$
\begin{equation}
\mathcal{P}_{j_{2k+2}\cdots j_{2n-1}}^{i_{2k+2}\cdots
i_{2n-1}}(R)=\sum_{s=0}^{n-1-k}\sum_{p=k}^{k+s}\Delta
^{(p)}(C_{k+s,p})_{j_{2k+2}\cdots j_{2n-1}}^{i_{2k+2}\cdots
i_{2n-1}}F_{j_{2n-2}j_{2n-1}}^{i_{2n-2}i_{2n-1}}\cdots
F_{j_{2n-2}j_{2n-1}}^{i_{2n-2}i_{2n-1}}\,,
\end{equation}
where $C_{k+s,p}$ are coefficients to be determined.  Evaluating the polynomial in the asymptotic region, we find
\begin{equation}
\mathcal{P}_{j_{2k+2}\cdots j_{2n-1}}^{i_{2k+2}\cdots i_{2n-1}}(-\ell _{%
\mathrm{eff}}^{-2}\delta ^{[2]})=\Delta
^{(k)}(C_{kk})_{j_{2k+2}\cdots j_{2n-1}}^{i_{2k+2}\cdots i_{2n-1}}
\end{equation}
where the coefficient turns to be
\begin{equation}
(C_{kk})_{j_{2k+2}\cdots j_{2n-1}}^{i_{2k+2}\cdots i_{2n-1}}=\frac{\left(
-1\right) ^{n-k-1}2^{2n-3}\left( n-1\right) !^{2}}{(2n-1)!}\,\ell _{\mathrm{%
eff}}^{2}\,{\delta }_{{j}_{2k+2}{j}_{2k+3}}^{{i}_{2k+2}{i}_{2k+3}}\cdots {%
\delta }_{{j}_{2n}{j}_{2n-1}}^{{i}_{2n}{i}_{2n-1}} \,.
\end{equation}
Taking the behavior of the extrinsic curvature (\ref{extrinsic}), bulk curvature (\ref{Riem}) and the relation between the AdS curvature with the Weyl tensor, we can evaluate the charge asymptotically obtaining
\begin{eqnarray}
\tau _{i}^{j} &=&\frac{{{\ell _{\mathrm{eff}}{(-1)}^{k}\Delta ^{(k)}}}}{%
16\pi G\,{2}^{n-1}(2n-2)!}\,\delta _{ii_{2}\cdots {i}_{2k}{i}_{2k+1}\dots {i}%
_{2n}i_{2n-1}}^{jj_{2}\cdots {j}_{2k}{j}_{2k+1}\dots {j}_{2n}j_{2n-1}}\,{W}_{%
{j}_{2}{j}_{3}}^{{i}_{2}{i}_{3}}\cdots {W}_{{j}_{2k}{j}_{2k+1}}^{{i}_{2k}{i}%
_{2k+1}}\times  \notag \\
&&\qquad \qquad \qquad\qquad \qquad \times {\delta }_{{j}_{2k+2}{j}_{2k+3}}^{%
{i}_{2k+2}{i}_{2k+3}}\cdots {\delta }_{{j}_{2n}{j}_{2n-1}}^{{i}_{2n}{i}%
_{2n-1}}\,.
\end{eqnarray}

Contracting the antisymmetric deltas, the charge density has similar form as in even dimensions, see eq.(\ref{teven}). Namely, it is polynomial of order $k$ in the Weyl tensor,
\begin{equation}
\tau ^{ j }_{ i }=\frac { { { \ell _{ \rm{ eff } }{ (-1) }^{ k }\Delta ^{ (k) }(2n-2k-1)! } } }{ 16\pi G\, { 2 }^{ k }(2n-2)! } \, \delta _{ ii_{ 2 }\cdots { i }_{ 2k }{ i }_{ 2k+1 } }^{ jj_{ 2 }\cdots { j }_{ 2k }{ j }_{ 2k+1 } }\, { W }_{ { j }_{ 2 }{ j }_{ 3 } }^{ { i }_{ 2 }{ i }_{ 3 } }\cdots { W }_{ { j }_{ 2k }{ j }_{ 2k+1 } }^{ { i }_{ 2k }{ i }_{ 2k+1 } } \label{todd}
\end{equation}
This last expression is valid for theories with degeneracy level in the interval $1\leq k \leq n-1$. Once again, it reduces to Conformal Mass when $k=1$  (\ref{q=E}). For $k=n$ (CS AdS gravity), the charge vanishes identically.


\chapter{Conclusions} 

\label{conclusions}
In this thesis, we have found a direct link between the multiplicity of a given \textit{branch} of Lovelock-AdS gravity and the non-linearity of its conserved charges.
Taking as a starting point a finite definition of conserved quantities, we have explicitly shown that they can be expanded in $q$ powers of the Weyl tensor. For a theory with degeneracy $k$, all the divergent orders of the expansion, i.e., $(Weyl)^{q}$ with $q<k$, vanish. Therefore, the leading part of the charge in $D$ dimensions contains a product of $k$ Weyl tensors 
\begin{equation}
\tau _{i}^{j}=\frac{{{{(-1)}^{k}}}\ell _{\mathrm{eff}}{{{\ }\Delta
^{(k)}(D-2k-2)!}}}{16\pi G\,{2}^{k}(D-3)!}\,\delta _{ii_{2}\cdots {i}_{2k}{i}%
_{2k+1}}^{jj_{2}\cdots {j}_{2k}{j}_{2k+1}}\,{W}_{{j}_{2}{j}_{3}}^{{i}_{2}{i}%
_{3}}\cdots {W}_{{j}_{2k}{j}_{2k+1}}^{{i}_{2k}{i}_{2k+1}}\,. \label{finalcharge}
\end{equation}
The degeneracy condition (\ref{DeltaK}) as an overall factor in the charges shows the validity of the formula for a particular gravity theory with multiplicity $k$. \\
The expression (\ref{finalcharge}) can be applied to calculate the mass in Lovelock Unique Vacuum theories. In the maximally degenerate case of even dimensions $D=2n$ ($k=n-1$), i.e., Born-Infeld AdS gravity, the charge is a product of $n-1$ Weyl tensors or AdS curvatures that is proportional to a linear expression in the total energy of the black hole \cite{Arenas-Henriquez:2019rph} and matches with the result in ref.\cite{Crisostomo:2000bb,Pang:2011cs, Miskovic:2007mg}.\\ 
On the other hand, $\tau_{i}^{j}$ is identically zero for Chern-Simons AdS gravity ($D=2n+1$). The last term in the charge expansion does not contribute and all the other orders vanishes due to the condition (\ref{DeltaK}). From the point of view of the black hole solution, the mass parameter does not falls off, and there is a mass gap with respect to global AdS \cite{Crisostomo:2000bb}. As the charge is always factorizable by the AdS curvature, the energy can not come from there. Instead, the mass for black holes in such a theory must come from $\tau_{(0)}$ what indicates a qualitative difference between CS AdS and theories with $k<n$. This is somewhat expected, as the mass term in $f(r)$ is a constant, can be interpreted as a massive state with respect to the one given by global AdS vacuum.   \\
Our result provides insight on holographic properties of degenerate theories as is the case of CS AdS gravity whose CFT dual description is of an unusual sort \cite{Banados:2005rz,Banados:2006fe, Grozdanov:2016fkt,Cvetkovic:2017fxa}.
In particular, the connection between the $C_{T}$ coefficient in holographic correlation functions and the first degeneracy condition has been recently emphasized in \cite{Bueno:2018yzo,Bueno:2018uoy,Mir:2019ecg}. Nevertheless, as this coefficient is related to the a-charge in the $A$-type anomaly \cite{Imbimbo:1999bj,Li:2018drw}, our results suggest that for a more general scenario, the computations may involve the contribution of higher degeneracy conditions.
\\
We are on the way to further understand holographic consequences of the above results.


\appendix 



\chapter{Generalized Kronecker Delta}

\label{AppendixA}
The Levi-Civita tensor in $D$ dimensions is completely antisymmetric tensor density defined by 
\begin{eqnarray*}
\varepsilon_{\mu \nu \alpha \beta \gamma \dots} = \left\{ \begin{matrix} +1 & \mbox{if }(\mu, \nu, \alpha, \beta, \gamma, \dots) \mbox{ is an even permutation of } (0,1,2,\dots,D-1) \\ -1 & \mbox{if }(\mu, \nu, \alpha, \beta, \gamma, \dots)  \mbox{ is an odd permutation of } (0,1,2,\dots,D-1)\\ 0 & \quad \quad \quad \mbox{if any two labels are the same.} \end{matrix}\right.
\end{eqnarray*}
The generalized anti-symmetric Kronecker delta of rank $D$ is totally antisymmetric tensor defined by
\begin{eqnarray}{\varepsilon^{\nu_{1}...\nu_{p}}\varepsilon_{\mu_{1}...\mu_{p}} = \delta^{[\nu_{1}...\nu_{D}]}_{[\mu_{1}...\mu_{D}]} := \begin{vmatrix}
\delta^{\nu_{1}}_{\mu_{1}} & \delta^{\nu_{2}}_{\mu_{1}} & \cdots & \delta^{\nu_{D}}_{\mu_{1}} \\ 
\delta^{\nu_{1}}_{\mu_{2}} & \delta^{\nu_{2}}_{\mu_{2}} & \cdots & \delta^{\nu_{D}}_{\mu_{2}} \\
\vdots & \ & \ddots &  \\
\delta^{\nu_{1}}_{\mu_{D}} & \delta^{\nu_{2}}_{\mu_{D}}  & \dots & \delta^{\nu_{D}}_{\mu_{D}}
\end{vmatrix} }     \,.
\end{eqnarray}
The contraction of $\mathit{k \leq D}$ indices of a generalized Kronecker delta of rank  $\mathit{p}$ and $k$ Kronecker symbols what gives a Kronecker delta of rank $\mathit{p-k}$,
\begin{eqnarray}{ \delta^{[\nu_{1}...\nu_{k}...\nu_{p}]}_{[\mu_{1}...\mu_{k}...\mu_{p}]}\delta^{\mu_{1}}_{\nu_{1}}\cdots\delta^{\mu_{k}}_{\nu_{k}}   = \frac{(D-p+k)!}{(D-p)!} \delta^{[\nu_{k+1}...\nu_{p}]}_{[\mu_{k+1}...\mu_{p}]} }\,, \label{identitydelta}
\end{eqnarray}
where $D$ is the range of the indices. \\
The explicit form of the Kronecker delta of rank 2 is
\begin{eqnarray}
{ \delta^{[\mu \nu]}_{[\alpha \beta]}} := \delta^{\mu \nu}_{\alpha \beta}= \delta^{\mu}_{\alpha}\delta^{\nu}_{\beta} - \delta^{\mu}_{\beta}\delta^{\nu}_{\alpha}
\end{eqnarray}


\chapter{Gauss-normal foliation}

\label{AppendixB} 
For a spacetime $\mathcal{M}$ endowed by a metric, the local coordinates $x^{\mu}=(r,x^{i})$ can always be chosen in the Gauss-normal form, such that the line element becomes
\begin{equation}
ds^2=N^2(r)\,dr^2+h_{ij}(r,x)\,dx^idx^j\,,
\end{equation}
where $h_{ij}$ is the induced metric at a fixed radial coordinate $r$.
This form of the metric is particularly useful to express the bulk quantities in terms of the boundary tensors.
In this frame, the extrinsic curvature is defined by the formula
\begin{equation}
K_{ij}=-\frac{1}{2}\mathcal{L}_nh_{ij}=-\frac{1}{2N} {\partial_r }h_{ij}\,, \label{extrinsicK}
\end{equation}
where $\mathcal{L}_n$ is the Lie derivative along a radial normal $n_{\mu}=N\delta_{\mu}^{r}$. The
foliation leads to the Gauss-Codazzi relations that are the components of the Riemann tensor decomposed in the given frame,
\begin{eqnarray}
R_{jl}^{ir}&=&\frac { 1 }{ N }\, ( {\nabla }_{ l }{\ K }_{ j }^{i }-{\nabla }_{ j }{K }_{ l }^{ i })\,,  \nonumber \\
R_{ jr }^{ ir }&=&\frac { 1 }{ N }\, {\partial }_{ r }{K }_{ j }^{ i}-{K }_{ n }^{ i }{K }_{ j }^{ n }\,,  \label{gausscodazzi} \\
{R }_{ jl }^{ ik }&=&{\mathcal{R }}_{ jl }^{ ik }(h)-{K }_{ j }^{ i }{K }_{ l }^{ k }+{K }_{ l }^{ i }{K }_{ j }^{ k }\,. \nonumber
\end{eqnarray}
Here, $\nabla_{j}=\nabla_{j}(h)$ is the covariant derivative defined with respect to the induced metric and ${\mathcal{R }}_{ jl }^{ ik}(h)$ is the intrinsic curvature of the boundary $\partial \mathcal{M}$. The indices on the right hand side of (\ref{gausscodazzi}) are lowered and rised by the induced metric and its inverse. 

\chapter{Properties of the Riemann and Weyl tensor}

\label{AppendixC} 
The Riemann tensor measures a deviation of a given metric from the flat space, and its components are given by
\begin{equation}
R^{\mu}_{\nu \alpha \beta}=\partial_{\alpha}\Gamma^{\mu}_{\beta \nu}  -\partial_{\beta}\Gamma^{\mu}_{\alpha \nu}+\Gamma^{\mu}_{\alpha \rho}\Gamma^{\rho}_{\beta \nu}-\Gamma^{\mu}_{\beta \rho}\Gamma^{\rho}_{\alpha \nu}\,.
\end{equation}
It has the following properties and symmetries,
\begin{align}
{}&R_{\mu \nu}= R^{\rho}_{\mu \rho \nu}\,,&\mbox{(Ricci tensor)}\\
{}&R^{\mu \rho}_{\alpha \beta}=g^{\rho \nu}R^{\mu}_{\nu \alpha \beta}\,,& \\
{}&R=R^{\mu \nu}_{\mu \nu}\,,& \mbox{(Ricci scalar)}\\
{}&R_{\mu \nu \alpha \beta}=R_{\alpha \beta \mu \nu}\,, & \\
{}&R_{\mu [\nu \alpha \beta]}=0\,,&\mbox{(First Bianchi identity)} \\
{}&R_{\mu \nu [\alpha \beta ;\rho]}=0.&\mbox{(Second Bianchi identity)}
\end{align}

For a static spherically symmetric black hole with metric (\ref{ansatz}) in the local coordinates $x^{\mu}=(t,r,y^{m})$, the only nonvanishing components of the Riemann tensor are

\begin{equation}
\begin{array}{ll}
R_{tr}^{tr}=-\frac{1}{2}\,f^{\prime \prime }\,,\medskip \qquad &
R_{rm}^{rn}=R_{tm}^{tn}=-\frac{f^{\prime }}{2r}\,\delta _{m}^{n}\,, \\
R_{m_{1}m_{2}}^{n_{1}n_{2}}=\frac{k-f}{r^{2}}\,\delta
_{m_{1}m_{2}}^{n_{1}n_{2}}\,. &
\end{array}
\end{equation}
On the other hand, the Weyl tensor is defined by
\begin{equation}
W_{\alpha \beta}^{ \mu \nu }=R_{ \alpha \beta }^{ \mu \nu }-
\frac { 1 }{ D-2 } \,\delta_{ [\alpha }^{ [\mu }R_{ \beta ] }^{ \nu] }
+\frac {R}{(D-1)(D-2) } \,{\delta }_{ \alpha \beta }^{\mu \nu }\,,
\end{equation}
where the antisymmetrization 
\begin{equation}
   \delta_{ [\alpha }^{ [\mu }R_{ \beta ] }^{ \nu] }=\delta_{\alpha}^{\mu}R_{\beta}^{\nu} - \delta_{\beta}^{\mu}R_{\alpha}^{\nu}+\delta_{\beta}^{\nu}R_{\alpha}^{\mu}-\delta_{\alpha}^{\nu}R_{\beta}^{\mu}\,.
\end{equation}
If we introduce the Schouten tensor 
\begin{equation}
    S^{\mu}_{\nu}=\frac{1}{D-2}\left(R^{\mu}_{\nu}-\frac{1}{2(D-1)}\delta^{\mu}_{\nu}R\right)\,,
\end{equation}
then the Weyl tensor can be expressed in terms of the Schouten and Riemann tensor as
\begin{equation}
    W_{\alpha \beta}^{ \mu \nu }=R_{ \alpha \beta }^{ \mu \nu }-S^{[\mu}_{[\alpha}\delta^{\nu]}_{\beta]}\,.
\end{equation}
Using the properties of the Riemann tensor, the Weyl tensor satisfies
\begin{align}
{}&W^{\mu \nu}_{\mu \nu}= 0\,, \\
{}&W^{\mu \alpha }_{\nu \alpha}=0\,,\\
{}&W_{\mu \nu \alpha \beta}= W_{[\mu \nu][\alpha \beta]}\,,\\
{}&W_{\mu  \nu \alpha \beta} =W_{\alpha \beta \mu \nu}\,,\\
{}&W_{\mu [\nu \alpha \beta]}=0.
\end{align}
We define the electric part of the Weyl tensor as 
 \begin{equation}
 \label{Eij}
 {E}_{i}^{j}=\frac{1}{D-3}\,{W}_{i\nu }^{j\mu }{n}_{\mu }{n}^{\nu }\,,
 \end{equation}
 where $x^{i}=(t,y^{m})$ are the coordinates on the boundary and ${n}_{\mu }$ is a spacelike normal to the boundary.
\chapter{Explicit factorization of the charge}

\label{AppendixD} 

For the sake of simplicity, we have used a shorthand notation where $\delta^{[k] }$ denotes the antisymmetric Kronecker
delta of rank $k$, that is, $\delta _{i_1\cdots i_k}^{j_1\cdots j_k}$.\\
In turn, $\delta ^{j[k]}_{i[k]}$ corresponds to a totally antisymmetric Kronecker delta of rank $(k+1)$ with the indices $i$ and $j$ fixed. This means that $\delta _{ii_1\cdots i_k}^{jj_1\cdots j_k}\equiv \delta ^{j[k]}_{i[k]} $.
\\
Additionally, we can redefine some parameters in order to make the computations simpler. 
For instance, the degeneracy condition (\ref{DeltaK}) can be expressed as%
\begin{equation}
\Delta ^{(q)}=\sum_{p=q}^{n-1}\binom{p}{q}\ell _{\mathrm{eff}}^{2(q-p)}\beta
_{p}\,,  \label{Delta-q}
\end{equation}%
where the coefficients $\beta _{p}$ includes the Lovelock couplings $%
\alpha _{p}$ and all the other dimensionless factors, 
\begin{equation}
\beta _{p}=\frac{(-1)^{p-1}(D-3)!\,\alpha _{p}}{(D-2p-1)!}\,.  \label{beta}
\end{equation}
We will extensively use the tensorial quantity \begin{equation}
F=A+R=\frac{1}{{\ell _{\mathrm{eff}}^{2}}}%
\,\delta ^{[2]}+R\,, \label{tensorAF}
\end{equation}
that is nothing but the AdS curvature defined in eq.(\ref{AdScurvature}).

\section{Even dimensions} \label{evenfactorization}
Consider  the charge (\ref{echarge}) written in the shorthand notation as
\begin{equation}
\tau^{i}_{j}= \frac{{{\ell _{\mathrm{eff}}^{2(n-1)}}}}{16\pi G\,{2}^{n-2}}\,\delta_{m}
^{\lbrack 2n-1]j}\,K_{i}^{m}\,f\,,  \label{qevenfact}
\end{equation}
where we introduced the auxiliary function

\begin{equation}
f(R,\delta^{[2]})=\sum_{p=1}^{n-1}{\frac{p{\alpha _{p}\ell _{\mathrm{eff}}^{2(1-p)}R^{p-1}}}{%
(2n-2p)!}}\left[ \left( \frac{1}{\ell _{\mathrm{eff}}^{2}}\,\delta ^{\lbrack
2]}\right) ^{n-p}-(-R)^{n-p}\right] .
\end{equation}%
The function $f$ can be written in terms of (\ref{beta}) and (\ref{tensorAF}) as
\begin{equation}
f=\frac{1}{2\left( 2n-3\right) !}\sum_{p=1}^{n-1}{\frac{p{\beta _{p}\ell _{%
\mathrm{eff}}^{2(1-p)}}}{n-p}\,}\left[ \left( A-{F}\right) {^{p-1}}%
A^{n-p}-\left( A-{F}\right) {^{n-1}}\right] . \label{ef1}
\end{equation}

We are interest in the Lovelock AdS gravities where the Lovelock coefficients $\alpha_{p}$ are such that the theory has a degenerate AdS vacuum of order $k$ given through the criterion (\ref{DeltaK}). Our main goal is to show that $f$ can be factorizable and cast to the form
\begin{equation}
f=(A+R)^{k}\mathcal{P}(R)\,.  \label{factor}
\end{equation}%
In that way, the problem is reduced to obtain the form of the polynomial $\mathcal{P}(R)$. To this end, we have to expand the binomials of the function (\ref{ef1}) in series 
\begin{eqnarray}
f &=&\frac{1}{2(2n-3)!}\sum_{p=1}^{n-1}{\frac{p\beta _{p}{\ell _{\mathrm{eff}%
}^{2(1-p)}}}{n-p}}\left( \sum_{q=1}^{p-1}\frac{\left( -1\right) ^{q}\left(
p-1\right) !}{q!\left( p-1-q\right) !}\right.   \notag \\
&&\qquad \qquad -\left. \sum_{q=1}^{n-1}\frac{\left( -1\right) ^{q}\left(
n-1\right) !}{q!\left( n-1-q\right) !}\right) F^{q}A^{n-1-q}\,, \label{fevenn}   
\end{eqnarray}%
whose first term, $q=0$ cancels identically giving a sum starting from $q=1$. Therefore, the power series takes the form
 \begin{equation}
f=\sum_{q=1}^{n-1}f_{q}\,F^{q}\,.  \label{f sum}
\end{equation}%

The coefficients $f_{q}$ are in fact a linear combination of degeneracy conditions $\Delta ^{(p)}$ in any dimension
\begin{equation}
f_{q}=\sum_{p=1}^{q}C_{qp}\,\Delta ^{(p)}\,,  \label{fq sum}
\end{equation}%
where $C_{qp}$ are constant tensorial coefficients. Nevertheless, this is not clear as first sight from the formula (\ref{fevenn}). Indeed, we can expand explicitly the first few terms of the series 
\begin{eqnarray}
f &=&\frac{1}{2(2n-3)!}\sum_{p=1}^{n-1}p\beta _{p}{\ell _{\mathrm{eff}%
}^{2(1-p)}}\left( FA^{n-2}-\frac{p+n-3}{2}\,F^{2}A^{n-3}\right.   \notag \\
&&+\left. \frac{p^{2}+np-6p+n^{2}-6n+11}{6}\,F^{3}A^{n-4}+\mathcal{O}%
(F^{4})\right) \,.
\end{eqnarray}%
and using the definition (\ref{Delta-q}), we can write
\begin{equation}
q!\Delta ^{(q)}\ell _{\mathrm{eff}}^{2(1-q)}=\sum_{p=q}^{n-1}p(p-1)\cdots
(p-q+1)\,\ell _{\mathrm{eff}}^{2(1-p)}\beta _{p}\,,
\end{equation}%
the auxiliary function becomes 
\begin{eqnarray*}
f &=&\frac{1}{2(2n-3)!}\left[ \Delta ^{(1)}FA^{n-2}-\left( \frac{n-2}{2}%
\,\Delta ^{(1)}+{\ell _{\mathrm{eff}}^{-2}\,}\Delta ^{(2)}\right)
\,F^{2}A^{n-3}\right.  \\
&&+\left. \left( \frac{n^{2}-5n+5}{6}\,\Delta ^{(1)}+\frac{n-3}{3}\,{\ell _{%
\mathrm{eff}}^{-2}\,}\Delta ^{(2)}+{\ell _{\mathrm{eff}}^{-4}\,}\Delta
^{(3)}\right) \,F^{3}A^{n-4}+\mathcal{O}(F^{4})\right] .
\end{eqnarray*}
This last expression possess the desire form \ref{fq sum}. For a given vacuum of degeneracy $q$, the only nonvanishing coefficient in the series will be $C_{qq}$. It can be shown that this coefficient corresponds to
\begin{equation}
C_{qq}=\frac{\left( -1\right) ^{q-1}\ell _{\mathrm{eff}}^{2(1-q)}{A^{n-1-q}}%
}{2\left( 2n-3\right) !}=\frac{\left( -1\right) ^{q-1}\delta ^{\lbrack 2]}{%
^{n-1-q}}}{2\left( 2n-3\right) !\ell _{\mathrm{eff}}^{2(n-2)}}\,.
\label{cqq}
\end{equation}

Then, evaluating in the density tensor (\ref{qevenfact}) with all the indices yields
\begin{eqnarray}
\tau _{i}^{j} &=&\frac{{{{(-1)}^{k}\ell _{\mathrm{eff}}}}\,{{\Delta ^{(k)}}}%
}{16\pi G\,{2}^{n-1}(2n-3)!}\,\delta _{ii_{2}\cdots {i}_{2k}{i}_{2k+1}\dots {%
i}_{2n}i_{2n-1}}^{jj_{2}\cdots {j}_{2k}{j}_{2k+1}\dots {j}_{2n}j_{2n-1}}{W}%
_{{j}_{2}{j}_{3}}^{{i}_{2}{i}_{3}}\cdots {W}_{{j}_{2k}{j}_{2k+1}}^{{i}_{2k}{i%
}_{2k+1}}\times   \notag \\
&&\qquad \qquad \qquad \qquad \qquad \times {\delta }_{{j}_{2k+2}{j}%
_{2k+3}}^{{i}_{2k+2}{i}_{2k+3}}\cdots {\delta }_{{j}_{2n}{j}_{2n-1}}^{{i}%
_{2n}{i}_{2n-1}}\,.
\end{eqnarray}%
Using the identity (\ref{identitydelta}) we can contract the deltas in the above expression

\begin{equation}
\delta _{ii_{2}\cdots {i}_{2k}{i}_{2k+1}\dots {i}_{2n}i_{2n-1}}^{jj_{2}%
\cdots {j}_{2k}{j}_{2k+1}\dots {j}_{2n}j_{2n-1}}{\delta }_{{j}_{2k+2}{j}%
_{2k+3}}^{{i}_{2k+2}{i}_{2k+3}}\cdots {\delta }_{{j}_{2n}{j}_{2n-1}}^{{i}%
_{2n}{i}_{2n-1}}=2^{n-k-1}\left( 2n-2-2k\right) !\,\delta _{ii_{2}\cdots {i}%
_{2k}{i}_{2k+1}}^{jj_{2}\cdots {j}_{2k}{j}_{2k+1}}\,,
\end{equation}%
arriving to
\begin{equation}
\tau _{i}^{j}=\frac{{{{(-1)}^{k}{{\ell _{\mathrm{eff}}}}\,{{\Delta ^{(k)}}}%
\left( 2n-2-2k\right) !}}}{16\pi G\,{2}^{k}(2n-3)!}\,\delta _{ii_{2}\cdots {i%
}_{2k}{i}_{2k+1}}^{jj_{2}\cdots {j}_{2k}{j}_{2k+1}}\,{W}_{{j}_{2}{j}_{3}}^{{i%
}_{2}{i}_{3}}\cdots {W}_{{j}_{2k}{j}_{2k+1}}^{{i}_{2k}{i}_{2k+1}}\,.
\label{Teven}
\end{equation}

\section{Odd dimensions}
Let us write the starting expression eq.(\ref{qodd}) written in terms of the auxiliary function $f$ as
\begin{equation}
\tau _{j}^{i}=\frac{(-1)^{n}(2n-1)}{16\pi G(2^{3n-4})(n-1)!^{2}}\,\delta ^{\lbrack
2n-1]j}\,K_{i}\,f,,  \label{qoddfactt}
\end{equation}%
where 
\begin{equation}
f=\sum_{p=1}^{n}\frac{p\beta _{p}\,\ell _{\mathrm{eff}}^{2(n-p)}}{2n-2p+1}%
\int\limits_{0}^{1}du\,\mathcal{I}_{p}(u)\,. \label{foddfact}
\end{equation}%
Here, $\mathcal{I}_{p}(u)$ corresponds to eq.(\ref{I}). Similarly as in the even dimensional case, we are going to expand $f$ as a power series in terms of $F$, as (\ref{f sum}). Using the redefinitions, the tensorial quantity $\mathcal{I}_{p}(u)$ acquires the form
\begin{equation}
\mathcal{I}_{p}=\left( F-A+u^{2}A\right)
^{n-1}-(u^{2}-1)^{n-1}A^{n-p}(A-F)^{p-1}\,.
\end{equation}%
If we expand the binomials $\left( F-A+u^{2}A\right) ^{n-1}$ and $\left( A-F\right) ^{p-1}$, the integrand in (\ref{foddfact}) becomes
\begin{equation}
\mathcal{I}_{p}=\left( \sum_{q=0}^{n-1}\frac{\left( n-1\right) !\left(
u^{2}-1\right) ^{n-1-q}}{q!\left( n-1-q\right) !}-\sum_{q=0}^{p-1}\frac{%
\left( -1\right) ^{q}\left( p-1\right) !(u^{2}-1)^{n-1}}{q!\left(
p-1-q\right) !}\right) A^{n-1-q}F^{q}\,.
\end{equation}%
We can drop the term $q=0$ which is cancelled identically. Therefore, the integral in $u$ can be done using the following identity
\begin{equation}
\int\limits_{0}^{1}du\,\left( u^{2}-1\right) ^{n-1}=\frac{\left( -1\right)
^{n-1}2^{2n-2}(n-1)!^{2}}{(2n-1)!}\,,
\end{equation}%
yielding to
\begin{eqnarray}
\int\limits_{0}^{1}du\,\mathcal{I}_{p} &=&\left( -1\right)
^{n-1}2^{2n-2}\left( n-1\right) !\left( \sum_{q=1}^{n-1}\frac{(n-q-1)!}{%
2^{2q}(2n-2q-1)!}\right.   \notag \\
&&-\left. \frac{(n-1)!}{(2n-1)!}\sum_{q=1}^{p-1}\frac{\left( -1\right)
^{q}\left( p-1\right) !}{\left( p-1-q\right) !}\right) \frac{\left(
-1\right) ^{q}}{q!}\,A^{n-1-q}F^{q}\,.
\end{eqnarray}%
Then, the auxiliary function takes the form of
\begin{eqnarray}
f &=&\left( -1\right) ^{n-1}2^{2n-2}\left( n-1\right) !\sum_{p=1}^{n}\frac{%
p\beta _{p}\,\ell _{\mathrm{eff}}^{2(n-p)}}{2n-2p+1}\,\left( \sum_{q=1}^{n-1}%
\frac{\left( -1\right) ^{q}(n-q-1)!}{2^{2q}q!(2n-2q-1)!}\right.   \notag \\
&&\qquad -\left. \frac{(n-1)!}{(2n-1)!}\sum_{q=1}^{p-1}\frac{\left(
-1\right) ^{q}\left( p-1\right) !}{q!\left( p-1-q\right) !}\right)
A^{n-1-q}F^{q}\,.  \label{fodd}
\end{eqnarray}

Proceeding in a similar way as in the even dimensional case, the coefficients in the expansion can be cast in such a way that they can be identified as the coefficients $f_{q}$ which are of the form (\ref{fq sum}).
A straightforward expansion of the above series gives the first few terms 
\begin{eqnarray}
f &=&\frac{\left( -1\right) ^{n}2^{2n-3}\left( n-1\right) !^{2}}{\left(
2n-1\right) !}\sum_{p=1}^{n}p\beta _{p}\,\ell _{\mathrm{eff}}^{2(n-p)}\left(
A^{n-2}F-\frac{2n+2p-5}{4}\,A^{n-3}F^{2}\right.   \notag \\
&&+\left. \frac{4p^{2}+4pn-22p+4n^{2}-20n+33}{24}\,A^{n-4}F^{3}+\mathcal{O}%
(F^{4})\right) \,.
\end{eqnarray}%
Writing the coefficients in terms of the degeneracy condition (\ref{Delta-q}), it gives
\begin{eqnarray}
f &=&\frac{\left( -1\right) ^{n}2^{2n-3}\left( n-1\right) !^{2}}{\left(
2n-1\right) !}\left[ \ell _{\mathrm{eff}}^{2(n-1)}\Delta
^{(1)}\,A^{n-2}F\right.   \notag \\
&&-\left( \ell _{\mathrm{eff}}^{2(n-2)}\Delta ^{(2)}+\frac{2n-3}{4}\,\ell _{%
\mathrm{eff}}^{2(n-1)}\Delta ^{(1)}\right) A^{n-3}F^{2}  \notag \\
&&+\left( \ell _{\mathrm{eff}}^{2(n-3)}\Delta ^{(3)}+\frac{2n-5}{6}\,\ell _{%
\mathrm{eff}}^{2(n-2)}\Delta ^{(2)}+\right.   \notag \\
&&\left. +\left. \frac{(2n-3)(2n-5)}{24}\,\ell _{\mathrm{eff}%
}^{2(n-1)}\Delta ^{(1)}\right) A^{n-4}F^{3}+\mathcal{O}(F^{4})\right] .
\end{eqnarray}%
As we have mentioned before, for a degenerate branch of order $q$, the only nonvanishing term is the coefficient $C_{qq}$ which is determined as
\begin{eqnarray}
C_{qq} &=&\frac{\left( -1\right) ^{n-q-1}2^{2n-3}\left( n-1\right) !^{2}}{%
(2n-1)!}\,\ell _{\mathrm{eff}}^{2(n-q)}A^{n-q-1}  \notag \\
&=&\frac{\left( -1\right) ^{n-q-1}2^{2n-3}\left( n-1\right) !^{2}}{(2n-1)!}%
\,\ell _{\mathrm{eff}}^{2}\,\delta ^{[2]n-q-1}\,. \label{cqq-odd}
\end{eqnarray}
Having this, the charge density (\ref{qoddfactt}) with all the indices becomes
\begin{eqnarray}
\tau _{j}^{i} &=&\frac{\left( -1\right) ^{k}\ell _{\mathrm{eff}}\,\Delta
^{(k)}}{16\pi G\,2^{n-1}(2n-2)!}\,\delta _{ii_{2}\cdots {i}_{2k}{i}%
_{2k+1}\dots {i}_{2n}i_{2n-1}}^{jj_{2}\cdots {j}_{2k}{j}_{2k+1}\dots {j}%
_{2n}j_{2n-1}}\,{W}_{{j}_{2}{j}_{3}}^{{i}_{2}{i}_{3}}\cdots {W}_{{j}_{2k}{j}%
_{2k+1}}^{{i}_{2k}{i}_{2k+1}}\times   \notag \\
&&\qquad \qquad \qquad \qquad \qquad \times {\delta }_{{j}_{2k+2}{j}%
_{2k+3}}^{{i}_{2k+2}{i}_{2k+3}}\cdots {\delta }_{{j}_{2n}{j}_{2n-1}}^{{i}%
_{2n}{i}_{2n-1}}\,.
\end{eqnarray}%
Once again, using the identity (\ref{identitydelta}), we can contract the deltas and find the final mass formula in odd dimensions
\begin{equation}
\tau _{j}^{i}=\frac{\left( -1\right) ^{k}\ell _{\mathrm{eff}}\,\Delta
^{(k)}\left( 2n-2k-1\right) !}{16\pi G\,2^{k}(2n-2)!}\,\delta _{ii_{2}\cdots
{i}_{2k}{i}_{2k+1}}^{jj_{2}\cdots {j}_{2k}{j}_{2k+1}}\,{W}_{{j}_{2}{j}_{3}}^{%
{i}_{2}{i}_{3}}\cdots {W}_{{j}_{2k}{j}_{2k+1}}^{{i}_{2k}{i}_{2k+1}}\,.
\label{Todd}
\end{equation}
Note that the maximal degree of $\tau_{i}^{j}$ in $F$ from the formula (\ref{fodd}) is of order $F^{n-1}$ and therefore, the charge density only contains coefficients $\Delta^{(q)}$ until $\Delta^{(n-1)}$ .  It implies that, in
the case of Chern-Simons AdS gravity, where all the coefficient $\Delta^{(q)}$ vanish, except $\Delta^{(n)}$, the charge will be
\begin{equation}
\tau_{\text{\textrm{CS,}}i}^{j}=0\,,\qquad k=n\,,
\end{equation}
and the energy of CS AdS black holes is fully contained in $\tau_{(0)i}^{j}$.
\chapter{Conformal mass in Lovelock-AdS gravity}

\label{AppendixE}
Conformal Mass is the expression for energy in AAdS spacetimes \cite{Ashtekar:1984zz,Ashtekar:1999jx}. It was constructed using the Penrose's conformal completation techniques where the conformal boundary is brought to the finite distance by means of a conformal transformation in order to define finite quantities in the spacetime.\\
When the AdS vacuum is non-degenerate ($k=1$), the charge density tensor is linear in the Weyl tensor,
\begin{equation}
\tau_{i}^{j} \sim \Delta'(\ell _{\mathrm{eff}}^{-2})
\delta_{ii_2i_3}^{jj_2j_3}\,{W_{j_2j_3}^{i_2i_3}}\,.
\end{equation}
 Expanding the contractions of the Weyl tensor,
\begin{eqnarray}
\delta _{ii_{2}i_{3}}^{jj_{2}j_{3}}\,W_{j_{2}j_{3}}^{i_{2}i_{3}}
&=&2\left(\delta _{i}^{j}W_{kl}^{kl}-2W_{ki}^{kj}\right)\,,
\end{eqnarray}
and using  the fact that the Weyl tensor is traceless, we can show that
\begin{equation}
W_{kl}^{kl} =0\,,\qquad W_{\mu i}^{\mu j}=W_{ri}^{rj}+W_{ki}^{kj} =0\,.
\end{equation}
In doing so, we find
\begin{eqnarray}
\delta _{ii_{2}i_{3}}^{jj_{2}j_{3}}\,W_{j_{2}j_{3}}^{i_{2}i_{3}}
&=&-4W_{ki}^{kj}=4W_{ri}^{rj}\,.  \label{deltaw}
\end{eqnarray}
By definition, the electric part of the Weyl tensor  in $D$ dimensions is defined by
\begin{equation}
E_i^j=\frac{1}{D-3}\,{W}_{ri}^{rj}\,,
\end{equation}
where we have replaced the following components of the normal to the boundary $n_{\mu }=(n_r,n_i)=(N,0)$.
Then, we obtain
\begin{equation}
\tau_i^j =-\frac{{\ell _{\mathrm{eff}}}}{8\pi G}\,\Delta'(\ell_{\mathrm{eff}}^{-2})\,E_{i}^{j}\,.
\label{q=E}
\end{equation}
which is the Conformal Mass in Lovelock-AdS gravity theories with a non-degenerate AdS vacuum.

\bibliographystyle{JHEP}
\bibliography{References.bib}


\end{document}